\documentclass[twocolumn,trackchanges]{aastex61}

\usepackage[normalem]{ulem}
\usepackage{footnote}
\usepackage{subfigure}
\usepackage{xspace}
\usepackage[utf8]{inputenc}
\usepackage{multirow,amsmath}
\usepackage{float}

\newcommand\msun{M\ensuremath{_{\odot}}\xspace}                   
\newcommand{\Msun}{\msun}                                         
\newcommand\mesh{\ensuremath{\delta_{\rm{mesh}}}\xspace}          

\newcommand{\code}[1]{\texttt{#1}}
\newcommand{\mesa}{\code{MESA}\xspace}
\newcommand{\MESA}{\mesa}

\begin{document}

\title{The Impact of Nuclear Reaction Rate Uncertainties \\ On The Evolution of Core-Collapse Supernova Progenitors}

\author[0000-0002-8925-057X]{C.~E.~Fields}
\altaffiliation{NSF Graduate Research Fellow, Ford Foundation Predoctoral Fellow}
\affiliation{Department of Physics and Astronomy, Michigan State University, East Lansing, MI 48824, USA}
\affiliation{Joint Institute for Nuclear Astrophysics - Center for the Evolution of the Elements, USA}

\author[0000-0002-0474-159X]{F.~X.~Timmes}
\affiliation{School of Earth and Space Exploration, Arizona State University, Tempe, AZ, USA}
\affiliation{Joint Institute for Nuclear Astrophysics - Center for the Evolution of the Elements, USA}

\author[0000-0003-3441-7624]{R.~Farmer}
\affiliation{Anton Pannenkoek Institute for Astronomy, University of Amsterdam, NL-1090 GE Amsterdam, the Netherlands}
\affiliation{School of Earth and Space Exploration, Arizona State University, Tempe, AZ, USA}

\author[0000-0003-2434-1128]{I.~Petermann}
\affiliation{School of Earth and Space Exploration, Arizona State University, Tempe, AZ, USA}
\affiliation{Joint Institute for Nuclear Astrophysics - Center for the Evolution of the Elements, USA}

\author[0000-0002-6828-0630]{William~M.~Wolf}
\affiliation{School of Earth and Space Exploration, Arizona State University, Tempe, AZ, USA}
\affiliation{Department of Physics, University of California, Santa Barbara, CA 93106, USA}

\author[0000-0002-5080-5996]{S.~M.~Couch}
\affiliation{Department of Physics and Astronomy, Michigan State University, East Lansing, MI 48824, USA}
\affiliation{Joint Institute for Nuclear Astrophysics - Center for the Evolution of the Elements, USA}

\email{fieldsc9@msu.edu}
\received{October 31, 2017}
\revised{December 6, 2017}
\accepted{December 16, 2017}

\shorttitle{Impact of Rate Uncertainties on SN Progenitors}
\shortauthors{Fields et al.} 

\begin{abstract}
We explore properties of core-collapse supernova progenitors with respect to
the composite uncertainties in the thermonuclear reaction rates by coupling
the reaction rate probability density functions provided by the STARLIB reaction rate
library with \MESA stellar models.
We evolve 1000 15\,\Msun models from the pre main-sequence to core O-depletion
at solar and subsolar metallicities for a total of 2000 Monte Carlo stellar models.
For each stellar model, we independently and simultaneously sample 665 thermonuclear
reaction rates and use them in a \MESA in situ reaction network that follows 127 isotopes from
$^{1}$H to $^{64}$Zn.
With this framework we survey the core mass, burning lifetime, composition, and structural properties
at five different evolutionary epochs.
At each epoch we measure the probability distribution function of the variations of each property
and calculate Spearman Rank-Order Correlation coefficients for each sampled reaction rate
to identify which reaction rate has the largest impact on the variations on each property.
We find that uncertainties in
$^{14}$N$(p,\gamma)^{15}$O,
triple-$\alpha$,
$^{12}$C$(\alpha,\gamma)^{16}$O,
$^{12}$C($^{12}$C,$p$)$^{23}$Na,
$^{12}$C($^{16}$O,$p$)$^{27}$Al,
$^{16}$O($^{16}$O,$n$)$^{31}$S,
$^{16}$O($^{16}$O,$p$)$^{31}$P, and
$^{16}$O($^{16}$O,$\alpha$)$^{28}$Si
reaction rates dominate the variations of the properties surveyed.
We find that variations induced by uncertainties in nuclear reaction rates 
grow with each passing phase of evolution, and at core
H-, He-depletion are of comparable magnitude to the variations 
induced by choices of mass resolution and network resolution. 
However, at core C-, Ne-, and O-depletion, the reaction rate uncertainties can dominate
the variation causing uncertainty in various properties of the stellar model in 
the evolution towards iron core-collapse.
\end{abstract}


\section{Introduction}
\label{sec:introduction}
Core-collapse supernova (SN) explosions are one possible fate of a star with a
zero age main-sequence mass of M\,$\gtrsim$ 9\,\Msun
\citep[e.g.,][]{woosley_2002_aa,woosley_2007_aa,farmer_2015_aa}.
The structure of the progenitor at the time
of explosion can lead to a large variety of observed transient phenomena
\citep[e.g.,][]{van-dyk_2000_aa,ofek_2014_aa,smith_2016_aa}.

For progenitors experiencing mass loss, stellar winds may strip the H-rich
envelope, and possibly some of the He-rich envelope, prior to core-collapse
\citep[e.g.,][]{smith_2014_ac,renzo_2017_aa}.
Explosions of these stars are characterized by an absence
of hydrogen absorption features and weak or non-existent absorption lines of
silicon in their spectra
\citep{smartt_2009_aa,dessart_2011_aa,smartt_2015_aa,reilly_2016_aa,sukhbold_2016_aa}.
Progenitors with most of the H-rich envelope present at the end of their life are
characterized as Type II supernovae that can be sub-divided into multiple classes
based on lightcurve and spectral properties
\citep{filippenko_1997_aa,wang_2008_ad,jerkstrand_2015_aa}.

In some cases, a massive star with sufficient rotational energy at
core collapse can produce a rapidly rotating, highly magnetic proto-neutron 
star capable of leading to a
significantly enhanced energetic transient. Such a scenario has been
postulated to explain the most energetic supernova observed to date,
ASASSN-15lh \citep{sukhbold_2016_ab,manos_2016_aa,chen_2016_aa},
although \citet{leloudas_2016_aa} offers on an alternative hypothesis
on the nature of ASASSN-15lh.

Alternatively, a massive star may undergo iron core-collapse
but the resulting shocks are insufficient to unbind the star, leading
to accretion onto the nascent proto-neutron star and pushing it past
its maximum mass. These ``failed supernovae'' \citep[e.g.,][]{oconnor_2011_aa}
can produce stellar mass black holes at the rate suggested by the detection of
GW150914, GW151226, and GW170104 
\citep{abbott_2016_aa,abbott_2016_ab,abbott_2017_aa},
although a broad consensus on which massive stars produce black holes
has not yet been reached
\citep{timmes_1996_ac, fryer_2001_ab, heger_2003_aa, eldridge_2004_aa,
zhang_2008_ab, ugliano_2012_aa, clausen_2015_aa, sukhbold_2016_aa,
muller_2016_ab, woosley_2016_aa, kruckow_2016_aa, sukhbold_2017_aa, limongi_2017_aa}.

For more massive progenitors, pair-instability leads to a partial collapse,
which in turn causes runaway burning in the carbon-oxygen core
\citep{fowler_1964_aa, rakavy_1967_ab, barkat_1967_aa, rakavy_1967_aa, fraley_1968_aa}.
A single energetic burst from nuclear burning can disrupt the entire star without
leaving a black hole remnant behind to produce a pair-instability supernova
\citep{ober_1983_aa, fryer_2001_aa, kasen_2011_aa, chatzopoulos_2013_ab}.  
Alternatively, a series of bursts can trigger a cyclic pattern of nuclear burning, 
expansion and contraction, leading to a pulsational pair-instability supernova
that leaves a black hole remnant
\citep{barkat_1967_aa, woosley_2007_aa, chatzopoulos_2012_aa, woosley_2017_aa, limongi_2017_aa}.
A variety of outcomes is possible depending on the star's mass and rotation.

At the heart of these evolutionary pathways
are nuclear reaction rates.
These rates regulate the evolution of the star and can
significantly modify the stellar structure of the progenitor star at the end of
its life. A direct consequence of uncertainties in the reaction rates can result in
differences in the nucleosynthesis and explosion properties
\citep{rauscher_2002_aa,woosley_2007_aa,sukhbold_2016_aa,rauscher_2016_aa}.

Most reaction rate libraries provide recommended nuclear reaction
rates based on experiment (when possible) or theory. Examples include
CF88 \citet{caughlan_1988_aa},
NACRE \citep{angulo_1999_aa,xu_2013_aa},
JINA REACLIB \citep{cyburt_2010_aa}, and
STARLIB \citep{sallaska_2013_aa}.
STARLIB takes the additional step of providing the median or recommended thermonuclear
reaction rate \emph{and} the factor uncertainty ($f.u.$) as a function of
temperature.  The factor uncertainty is an estimate of the uncertainty
associated with a reaction rate at a given temperature given the
available nuclear physics data.
Monte Carlo \citep{longland_2010_aa, longland_2012_aa, iliadis_2015_aa,iliadis_2016_aa}
or Bayesian \citep{iliadis_2016_aa,gomez-inesta_2017_aa} based
reaction rates generate probability density functions (PDFs) to provide a final
\emph{median} rate and a temperature-\emph{dependent} uncertainty.
The availability of formally derived temperature-dependent uncertainties
allows statistically rigorous studies on the impact of the composite uncertainty
on stellar models.

Reaction rate sensitivity studies have been considered for X-ray burst models
\citep{cyburt_2016_aa} and massive star models through core He-burning
\citep{west_2013_aa} and for $s$-process nucleosynthesis \citep{nishimura_2017_aa}.  
In some of these and similar studies,
temperature-\emph{independent} estimates of the reaction rate
uncertainties are applied as constant multiplicative factors on the recommended
rate at all temperatures. This method can lead to an under- or over-estimate
of the reaction rate for different stellar
temperatures. Another common approximation is ``post-processing'' of
thermodynamic trajectories from stellar models
\citep[e.g.,][]{magkotsios_2010_aa, rauscher_2016_aa, harris_2017_aa}, which also
usually use a constant multiplicative factor at all temperature points.
Post-processing thermodynamic trajectories neglect the feedback of the changes
in the reaction rates on the underlying stellar model.

\citealt{fields_2016_aa} (Paper F16) addresses some of the shortcomings of these
approximations by using a Monte Carlo stellar model framework with
temperature-dependent uncertainties on the reaction rates from
STARLIB.  Specifically, used on 3\,\Msun stellar models evolved from
the pre main-sequence to the first thermal pulse. Each of the 1000 models
uses one set of reaction rates generated from the reaction rate PDFs.
These Monte Carlo stellar models probed the effect of reaction rate
uncertainties on the structure and evolution of stars that form
carbon-oxygen (CO) white dwarfs. Paper F16 sample
26 reaction rates of the 405 total rates in the chosen reaction network, 
which can bias identifying the reactions that play role in altering the stellar structure.

In this paper, we apply the same Monte Carlo framework to massive star
models.  We consider all forward reactions in a suitable reaction
network (reverse rates are calculated by detailed balance) 
to eliminate potential biases from selecting a limited set of
reactions.  Our workflow couples temperature-dependent reaction rate
uncertainties from STARLIB \citep{sallaska_2013_aa} with Modules for 
Experiments in Stellar Astrophysics (\MESA) stellar models
\citep{paxton_2011_aa,paxton_2013_aa,paxton_2015_aa}.
We sample the reaction rates independently and
simultaneously according to their respective PDFs.
These sampled rates form input for 15\,\Msun models evolved from the
pre main-sequence to core O-depletion. We focus on 15\,\Msun models as they
approximately represent the most numerous SNe by number for a Salpeter
initial mass function with slope $\Gamma$\,=\,$-1.35$, and a lower limit of
9\,\Msun for stars that become SNe
\citep{salpeter_1955_aa,scalo_1986_aa,sukhbold_2014_aa,farmer_2015_aa}.
We consider solar and subsolar metallicities to explore the effect of reaction 
rate uncertainties on stars in different galactic environments.

This paper is novel in two ways. First, we sample a large number of
reaction rates (665 forward reactions) in a Monte Carlo stellar
model framework where the rates are sampled \emph{before} the
stellar model is evolved. This accounts for changes in the stellar structure
due to reaction rate uncertainties, and is fundamentally different than
post-processing schemes. Second, we quantify the variation
of key quantities of the stellar models at five key evolutionary
epochs. This allows determination of (1) the most important reactions overall, and
(2) when these key reactions play a crucial role in the life of
a massive star.  In short, this paper presents the first Monte Carlo
stellar evolution studies of massive stars that use PDFs for the nuclear 
reaction rate uncertainties and complete stellar models.

In Section~\ref{sec:input_physics} we describe the input physics of our models.
In Section~\ref{sec:sampling} we discuss our Monte Carlo stellar model
framework and quantify the uncertainty of a few key nuclear reactions. 
Before presenting the results of our
survey, we describe the characteristics of baseline
15\,\Msun models evolved using \emph{median} reaction
rates from STARLIB in Section~\ref{sec:baseline}.
In Section~\ref{sec:mcstars} we present our main results.
In Section~\ref{sec:discussion} we compare our results
to previous efforts and make an assessment of the overall impact
of the uncertainties due to nuclear reactions relative to
other quantified sources of uncertainty \citep[e.g.,][]{farmer_2016_aa}.
In Section~\ref{sec:summary} we summarize our results.

\section{Input Physics}
\label{sec:input_physics}

We evolve 15\,\msun models using \texttt{MESA} \citep[version
7624,][]{paxton_2011_aa,paxton_2013_aa,paxton_2015_aa}.  All models
begin with an initial metallicity of $Z=Z_{\odot}=0.0153$
\citep[``solar'',][]{caffau_2010_aa,grevesse_1998_aa,asplund_2009_aa,vagnozzi_2017_aa}
or $Z$=2$\times$10$^{-3}$$Z_{\odot}$=0.0003 (``subsolar'').  Solar
metallicity models use isotopic distributions from
\citet{lodders_2009_aa}, while subsolar models use the methods of
\cite{west_2013_ab}\footnote{Available from \url{http://mesa-web.asu.edu/gce.html}}.
The metallicity-dependent isotopic distributions from
\cite{west_2013_ab} reproduce $\alpha$ enhancement trends for a large
sample of low $Z$ stars in the Milky Way halo \citep{frebel_2010_aa}
thus motivating our choice for these distributions over solar-scaled
compositions.

\begin{figure}[!htb]
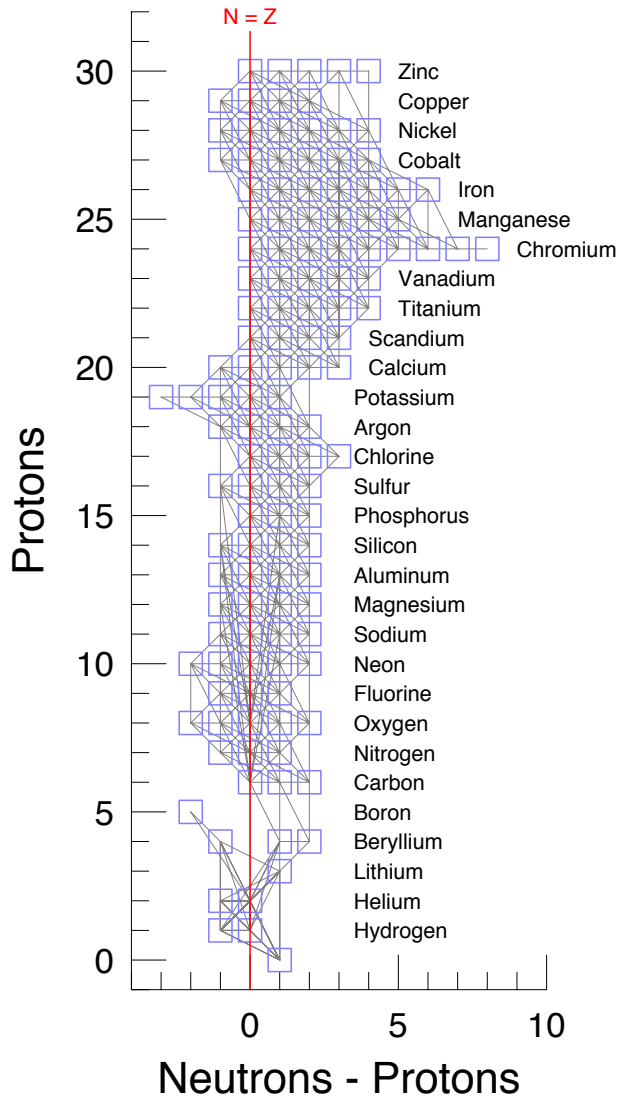

\centering{\includegraphics[width=1.0\columnwidth]{{{xz_plane_mesa_127_net_new}}}}
\caption{
Proton number versus neutron excess for the adopted 127 isotope reaction network.
Thermonuclear and weak reaction rates coupling the isotopes are marked by gray lines,
and symmetric matter ($N$=$Z$) is marked by a red line.}
\label{fig:nz_plane}
\end{figure}

\citet{farmer_2016_aa} show convergence of key quantities in 15\,\Msun \MESA\ models
at the $\simeq$\,10\% level when the reaction network contains $\gtrsim$ 127 isotopes.
Following their results, each stellar model utilizes the in-situ nuclear reaction
network \texttt{mesa\_127.net}, which follows 127 isotopes from
$^{1}$H to $^{64}$Zn coupled by 1201 reactions.
Figure~\ref{fig:nz_plane} shows the 127 isotopes and their
linking nuclear reactions. The isotopic abundance distributions
we use contains 288 isotopes from $^{1}$H to $^{238}$U. We add the residual
mass fraction ($\lesssim 10^{-5}$) of the 161 isotopes not in the reaction network
to the initial $^{1}$H mass fraction to maintain baryon number
conservation $\sum^{127}_{i=1} X_i=1$, where $X_i$ is the mass 
fraction of isotope $i$.

We include mass loss using
the \texttt{Dutch} wind loss scheme \citep{nieuwenhuijzen_1990_aa,
nugis_2000_aa,vink_2001_ab,glebbeek_2009_aa} with an efficiency of
$\eta$=0.8. We neglect the effects of rotation, magnetic fields, and rotation
induced mass loss in this study.

We use the Ledoux criterion for convection with an
efficiency parameter of $\alpha_{\rm{MLT}}=2.0$, and the
\texttt{mlt++} approximation for convection \citep{paxton_2013_aa}.  We include
convective boundary mixing (overshoot, thermohaline, and
semi-convection) with baseline values following \citet{farmer_2016_aa}.
For convective overshoot we use \mbox{ $f$ = 0.004 and   $f_{\rm{0}}=0.001$},
which can reproduce mass entrainment rates found in
idealized 3D simulations of explosive O-shell burning in massive stars
\citep{jones_2017_aa}.  For simplicity, we apply the same overshoot
efficiency to all boundaries.  For thermohaline mixing, we use
$\alpha_{\rm{th}}$ = 2.0 \citep{traxler_2011_aa,brown_2013_aa,garaud_2015_aa}.
Semi-convection uses an efficiency of $\alpha_{\rm{sc}}$ = 0.01
\citep{zaussinger_2013_aa,spruit_2013_aa}.

We use the \MESA control \texttt{mesh\char`_delta\char`_coeff}, \mesh,
to monitor mass resolution, which accounts for the
gradients in the structure quantities to decide whether a cell
should be split or merged. The default \MESA value is unity. In this
work, we use \mesh=0.5.  This results
in $\simeq$\,2300 cells at the terminal age main-sequence (TAMS),
$\simeq$\,4700 at core He-depletion, and $\simeq$\,2100 cells during
core O-burning. Section~\ref{sec:baseline} discusses the sensitivity of
our results to mass resolution.

We use several of \MESA's timestep controls.  The
parameter \texttt{varcontrol\char`_target}, $w_{\rm{t}}$, broadly controls
the temporal resolution by restricting the allowed relative variation
in the structure between timesteps.  The default value is
$w_{\rm{t}}$=$1\times10^{-4}$. In this work, we use
$w_{\rm{t}}$=$5\times10^{-5}$, except
during off-center C-burning where we use
$w_{\rm{t}}$=$1\times10^{-5}$ to further improve time resolution.
We also control the rate of fuel
depletion with the \texttt{delta\char`_lg\char`_X*} timestep controls, where
the asterisk denotes a major fuel (i.e. H, He, C, Ne, or O). In total,
we observe timesteps of $\Delta t \simeq
2\times10^{4}$ yr on the main sequence, $\Delta t \simeq
4\times10^{3}$ yr during core He-burning, and $\Delta t \simeq 12$ hr
during core O-burning.
Section~\ref{sec:baseline} discusses the sensitivity of
our results to temporal resolution.

For each stellar model, we sample 665 forward reaction rates from
STARLIB Archived Version 5 \citep{sallaska_2013_aa} simultaneously and
independently within their temperature-dependent uncertainties.  We
calculate reverse rates directly from the forward rates using detailed
balance. We utilize the work of \citet{alastuey_1978_aa} and
\citet{itoh_1979_aa} for reaction rate screening factors.  The fitting formula of
\citet{itoh_1996_aa} provide the thermal neutrino energy losses.  Weak
reactions rates, in order of precedence, are from
\citet{langanke_2000_aa}, \citet{oda_1994_aa}, and \citet{fuller_1985_aa}.

Each stellar model evolves from the pre main-sequence until the
central $X(^{16}$O) $\simeq$~1$\times$10$^{-3}$.
We use 1000 solar and subsolar stellar models, for a total
of 2000 Monte Carlo stellar models. All \texttt{MESA} inlists and 
many of the stellar models are available at {\url{http://mesastar.org}}.

\vspace{2cm}

\section{Reaction Rate Sampling}
\label{sec:sampling}
We construct a sampled nuclear reaction rate following
\citet{iliadis_2015_aa}.
We summarize the key characteristics here.
The STARLIB rate library provides the \emph{median} reaction rate,
$\left<\sigma v\right>_{\rm{med}}$, and the associated $f.u.$, over
the temperature range $10^{6-10}$ K. A log-normal PDF
is assumed for all reaction and decay rates,
and these PDFs are described by the location
and spread parameters, $\mu$ and $\sigma$, respectively.
These parameters are obtained using the median rate and $f.u.$
tabulated in STARLIB as
$\sigma = \textup{ln}~f.u.$ and
$\mu = \textup{ln}~\left<\sigma v\right>_{\rm{med}}$.
These two parameters give a complete description of the reaction rate
probability density at any temperature point and form the basis of our
sampling scheme.

A sampled reaction rate is drawn
from a log-normal distribution \citep[e.g.,][]{evans_2000_aa}
for an arbitrary quantity, $x$, as
\begin{equation}
x_i = e^{\mu + \sigma p_i } \equiv e^{\mu}(e^{\sigma})^{p_i}~.
\end{equation}
Using the relations for $\mu$ and $\sigma$, we obtain a sampled rate
distribution as a function of temperature from
\begin{equation}
\left<\sigma v\right>_{\rm{samp}}
= e^{\mu}(e^{\sigma})^{p_{i,j}}
= \left<\sigma v\right>_{\rm{med}} f.u.^{p_{i,j}}~,
\label{eq:sample_lambda}
\end{equation}
where $p_{i,j}$ is a standard Gaussian deviate with mean of zero and
standard deviation of unity. The $i$ index correspond to the stellar model 
of grid size $N$ and the $j$ index corresponds to the number of reactions 
sampled.

We refer to $p_{i,j}$ as the rate variation factor for the $j$-th reaction.
From Equation~\ref{eq:sample_lambda}, a rate variation
factor of $p_{i,j}=0$ corresponds to the \emph{median}
STARLIB reaction rate. For large rate variation factors, the
extent of change of the reaction rate at a given temperature point is
limited by the factor uncertainty.

For example, for the
$^{12}\rm{C}(\alpha,\gamma)^{16}\rm{O}$ reaction rate \citep{kunz_2002_aa},
STARLIB shows that the largest value of
factor uncertainty is $f.u.=1.403$ at $T=0.4$ GK. For typical extrema
of a Gaussian distribution such as those used to generate our rate
variation factors, one could expect values of $p_{i,j}=+3.5,-3.5$. In
such a scenario, this would represent a change in the sampled nuclear
reaction rate of $\left<\sigma v\right>_{\rm{samp}}$ $\simeq
3.27 \times \left<\sigma v\right>_{\rm{med}}$ for $p_{i,j}=+3.5$ and $
\simeq 0.31 \times \left<\sigma v\right>_{\rm{med}}$ for
$p_{i,j}=-3.5$ at $T=0.4~\textup{GK}$.  At all other temperature
points, the modification of the median rate may be less for the same
value of $p_{i,j}$.

\begin{figure}[!htb]
\centering
\includegraphics[width=\columnwidth]{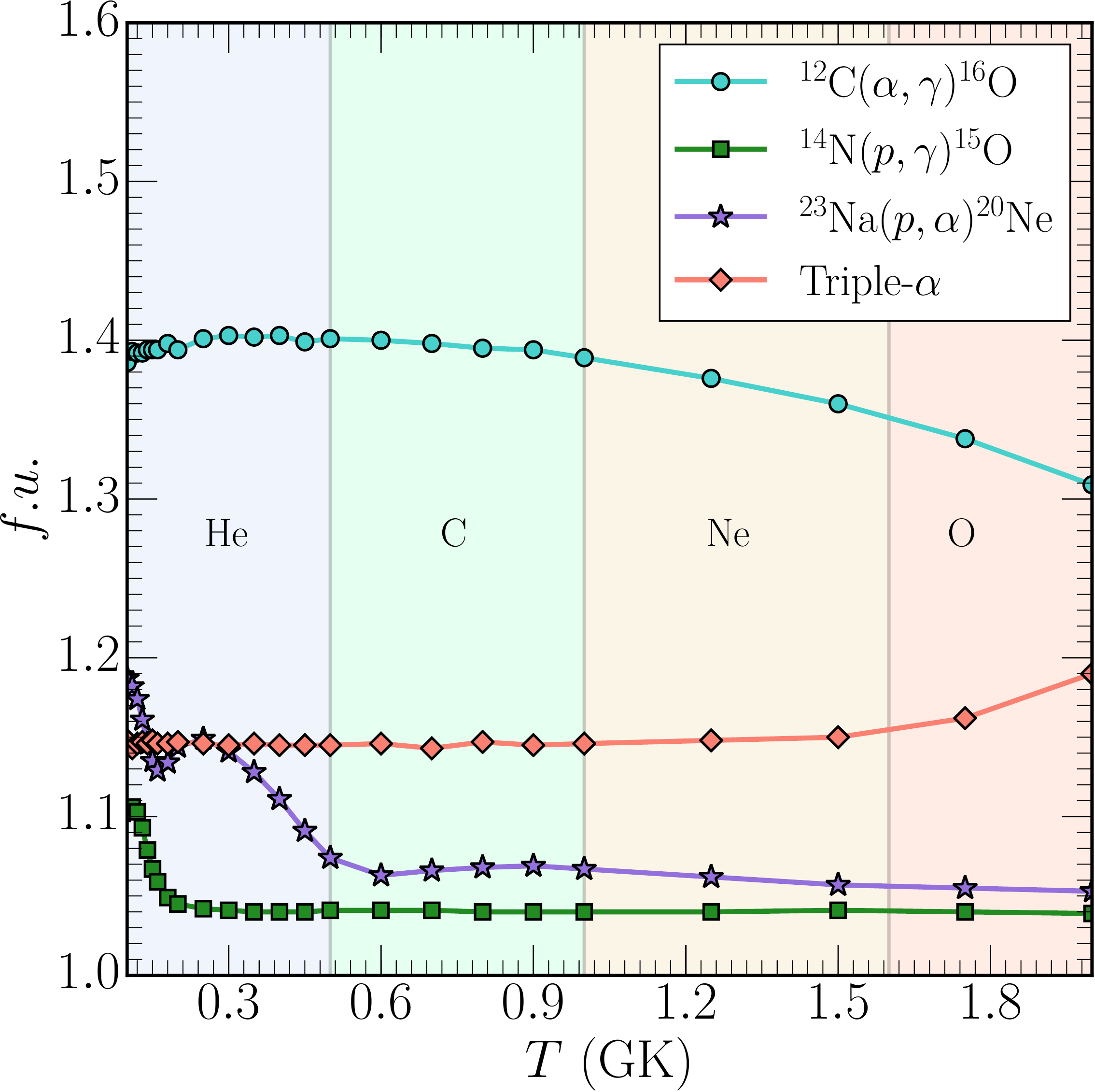}
\caption{The factor uncertainty as a function of temperature provided
by STARLIB for
the $^{12}$C($\alpha$,$\gamma$)$^{16}$O, $^{14}$N($p$,$\gamma$)$^{15}$O,
$^{23}$Na($p$,$\gamma$)$^{20}$Ne, and triple-$\alpha$ reactions
over different approximate core burning temperatures.
\label{fig:fu_factors}}
\end{figure}

In Figure~\ref{fig:fu_factors} we plot the $f.u.$
for the $^{12}$C($\alpha$,$\gamma$)$^{16}$O,
$^{14}$N($p$,$\gamma$)$^{15}$O, $^{23}$Na($p$,$\gamma$)$^{20}$Ne, and
triple-$\alpha$ reaction rates over typical core He-, C-, Ne-, and
O-burning temperatures.  The $^{12}$C($\alpha$,$\gamma$)$^{16}$O rate
has the largest factor uncertainty across the temperature ranges considered.
At higher temperatures such as those expected in more advanced burning stages
post core O-burning, the uncertainty in the $^{12}$C($\alpha$,$\gamma$)$^{16}$O
begins to be overtaken by the uncertainty in the triple-$\alpha$ reaction.

We simultaneously and independently sample 665 forward thermonuclear
reaction rates.  For each reaction, we generate $N$=1000 random Gaussian 
deviates to modifying the reaction rates in the stellar models.  Our choice for the
sample size is motivated by the scaling of the sampling error for perfectly 
uncorrelated distributions.  For such a distribution we expect a standard error of
$\sigma/\sqrt{N}\simeq3\%$. Since \MESA calculates
inverse rates directly from the forward rates using detailed balance,
we also implicitly sample the corresponding 665 inverse rates.
However, the corresponding inverse sampled rates are not independent
of the forward sampled reactions.

Reaction rates derived from Monte Carlo sampling of
experimental nuclear data are available for 33 of the 665 reactions
considered \citep{iliadis_2010_ab,sallaska_2013_aa,iliadis_2015_aa,iliadis_2016_aa}.
For other reactions, Monte Carlo or Bayesian derived rate
distributions are not yet available. In these such cases,
\emph{median} rate values and the corresponding temperature dependent
$f.u.$ are obtained from estimates of experimental uncertainty where
available.  In the absence of experimental nuclear physics input,
theoretical \emph{median} reaction rates are obtained from
Hauser-Feshbach model calculations with the \texttt{TALYS} software instrument
\citep{goriely_2008_aa}. Such theoretical rates are given a
constant uncertainty of $f.u.=10$ at all temperature points.

We assume the random Gaussian deviate is independent of
temperature, $p_{i,j}(T) =$ constant \citep{iliadis_2015_aa}.  This
simplification obtains similar levels of uncertainties
as more intricate sampling schemes
\citep{longland_2012_aa}.  We stress that despite this
simplification, the $f.u.$ provided by
STARLIB is temperature-\emph{dependent}. This allows us to
follow changes in the uncertainty that may occur due to different
resonance contributions.

\begin{figure}[!htb]
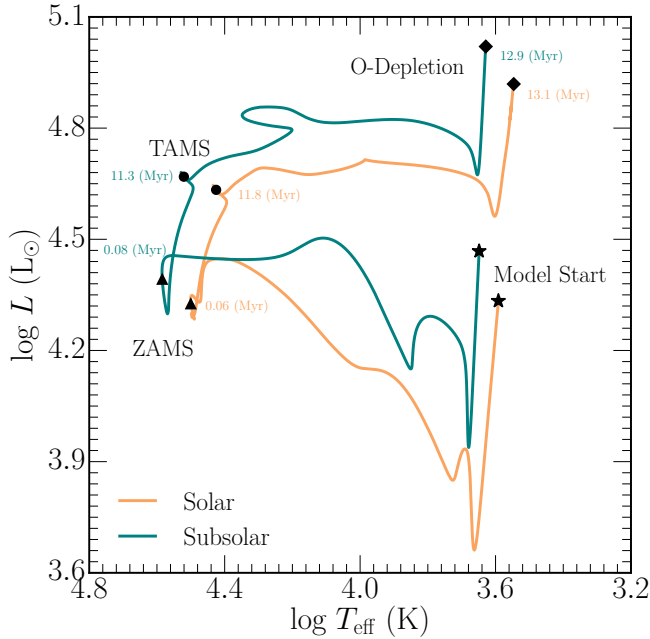

\centering{\includegraphics[width=1.0\columnwidth]{{{HR_eq_asp-cropped}}}}
\caption{
Hertzsprung-Russell diagram of the baseline 15\,\Msun solar and subsolar
models. Star symbols denote the beginning of each stellar model,
triangles denote the ZAMS, circles denote the TAMS, and diamonds
denote core O-depletion. Ages at these stages are annotated.}
\label{fig:baseline_HR}
\end{figure}

The sampled reaction rate distributions are then constructed
using Equation~(\ref{eq:sample_lambda}).
Each nuclear reaction rate in STARLIB has a total of 60
$T, \left<\sigma v\right>_{\textup{med}}$, and $f.u.$ data points.
A sampled reaction rate also contains 60 data points and is
then passed to \MESA in tabular form. \MESA interpolates between
data points to construct a smoothed sampled nuclear reaction rate 
defined by 10,000 reaction rate data points as a function of $T$.

\section{Properties of the baseline 15\,\Msun stellar models}
\label{sec:baseline}

Before presenting the results of our Monte Carlo stellar models survey,
we discuss the properties of the baseline 15\,\Msun solar and subsolar
models. These baseline models were evolved using the input physics 
described in Section~\ref{sec:input_physics} and the \emph{median} 
STARLIB nuclear reaction rates. A \emph{median} reaction rate is 
obtained in our sampling scheme by a Gaussian deviate of zero, 
$p_{i,j}=0$.

Figure~\ref{fig:baseline_HR} shows a Hertzsprung-Russell diagram
of the solar and subsolar baseline 15\,\Msun stellar models. The
start of the stellar models, the zero age main sequence (ZAMS),
terminal age main sequence, and the ending point of core 
O-depletion are annotated.  The subsolar model is brighter and hotter
than the solar model primarily because a smaller metallicity decreases
the opacity in the stellar atmosphere.  At ZAMS, the subsolar model has 
a luminosity and effective temperature of
$\log$($L$/L$_{\odot}$)~$\simeq$~4.39 and $\log$($T_{\rm{eff}}$/K)$\simeq$4.58
while the solar model has $\log$($L$/L$_{\odot}$)~$\simeq$~4.33 and
$\log$($T_{\rm{eff}}$/K)~$\simeq$~4.50.  The solar model spends
$\simeq$~11.2~Myr on the main sequence while the subsolar model spends
$\simeq$~11.7~Myr.

\begin{figure}[!htb]
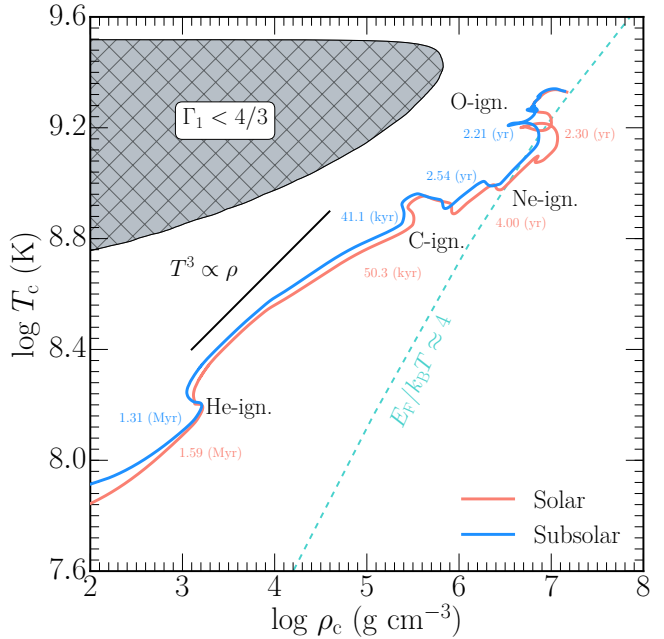

\centering{\includegraphics[width=1.0\columnwidth]{{{TRHo-cropped}}}}
\caption{
Evolution of the central density and temperature for the solar
and subsolar baseline models. Approximate core burning locations,
times until O-depletion, radiation entropy scaling relation - 
$T^{3}\propto \rho$, electron degeneracy line $E_{\rm{F}}/k_{\rm{B}}T\simeq4$
where $E_{\rm{F}}$ is the Fermi energy, and electron-positron 
pair dominated region are annotated.}
\label{fig:baseline_TRho}
\end{figure}

The ZAMS homology relations for CNO burning,
constant electron scattering opacity, and radiative transport
\citep{hoyle_1942_aa, faulkner_1967_aa,
pagel_1998_aa, bromm_2001_aa,portinari_2010_aa} are:
\begin{equation}
\begin{split}
\left ( \frac{T_{\rm{eff}}}{3 \times 10^4 \ {\rm K}} \right )
& \simeq
\left ( \frac{Z}{Z_{\odot}} \right )^{-1/20} \left ( \frac{M}{15 \ M_{\odot}} \right) ^{1/40} \\
\left ( \frac{R}{6 \ R_{\odot}} \right )
& \simeq
\left ( \frac{Z}{Z_{\odot}} \right)^{1/11} \left ( \frac{M}{15 \ M_{\odot}} \right) ^{5/11} \\
\left( \frac{L}{2 \times 10^4 L_{\odot}} \right )
& \simeq
\left ( \frac{Z}{Z_{\odot}} \right )^{-1/55} \left ( \frac{M}{15 \ M_{\odot}} \right ).
\end{split}
\label{eq:homologyrelations}
\end{equation}
The ZAMS positions of the solar and subsolar models in
Figure~\ref{fig:baseline_HR} and commensurate with the 
trends of Eq.~\ref{eq:homologyrelations}.

\begin{figure*}[!htb]
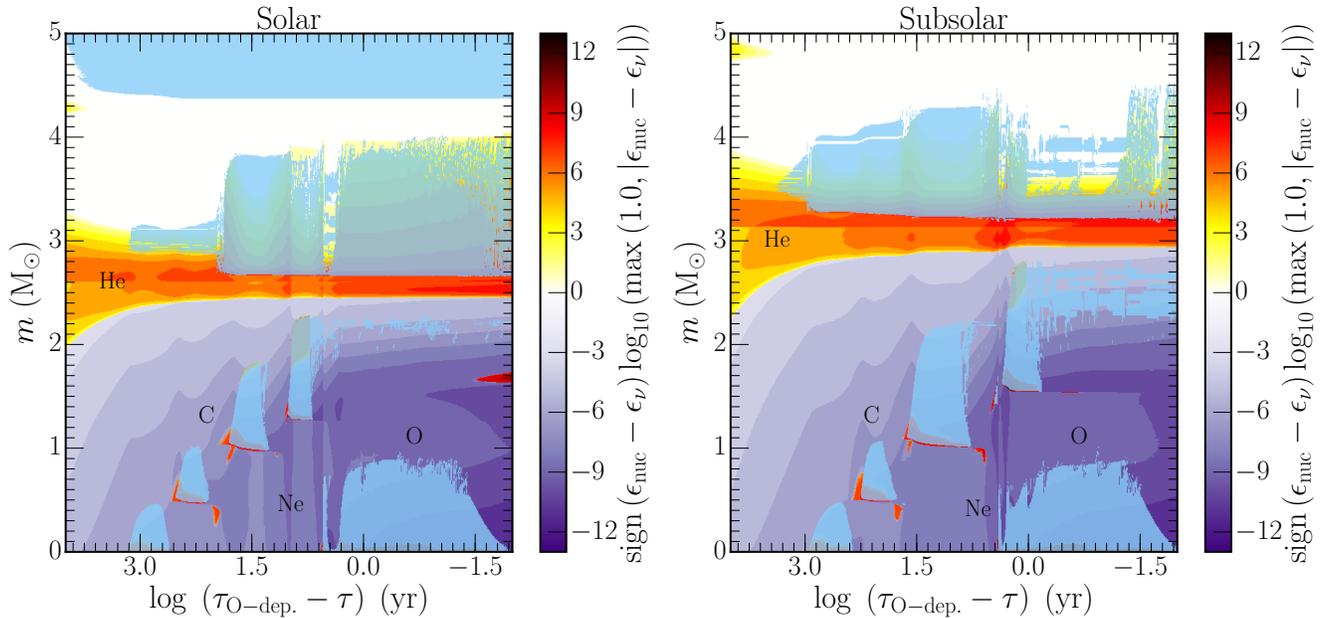

\begin{subfigure}
\centering{\includegraphics[width=\columnwidth]{{{solar_kipp-cropped}}}}
\end{subfigure}
\begin{subfigure}
\centering{\includegraphics[width=\columnwidth]{{{subsolar_kipp-cropped}}}}
\end{subfigure}
\caption{
Kippenhahn diagrams for the solar (left) and subsolar (right) baseline stellar models
post core He-burning. Annotated is the He-burning shell and the convective C, Ne,
and O-burning episodes. The x-axis is the logarithmic difference between
the age at O-depletion, $\tau_{\rm{O-dep.}}$, and the current age of the model, $\tau$.
Dark orange to red correspond to regions of strong nuclear burning, light to dark
purple to cooling regions, and white to regions balancing heating and cooling.
Blue shows convective regions, gray marks regions of convective overshoot.
Semi-convective and thermohaline regions are not shown.}
\label{fig:baseline_kipps}
\end{figure*}

At the TAMS, the nascent He-rich core is surrounded by
a thin H-burning shell. The core contracts and its temperature
increases, while the outer layers of the star expand and cool. The
star becomes a red giant
\citep[e.g.,][]{iben_1966_aa,iben_1991_aa,stancliffe_2009_aa,karakas_2014_aa}.
The solar model spends $\simeq$\,1.54~Myr undergoing convective core
He-burning and the subsolar model spends $\simeq$\,1.27~Myr.  At
He-depletion, the solar model has a He-core mass of $M_{\rm{He-Core}}$
$\simeq$\,4.24\,\Msun and a $^{12}$C/$^{16}$O ratio of 0.34. The
subsolar model has a more massive, slightly more C-rich core with
$M_{\rm{He-Core}}$~$\simeq$\,4.80\,\Msun and $^{12}$C/$^{16}$O
$\simeq$\,0.36.

The trajectory of the baseline models in the $T-\rho$ plane are shown in
Figure~\ref{fig:baseline_TRho}. In general, the tracks
are qualitatively similar. The largest difference is the subsolar model undergoes
hotter, less dense core burning. This is a result of the
decreased stellar envelope opacity and larger luminosity shown in
Figure~\ref{fig:baseline_HR}. Figure~\ref{fig:baseline_kipps} shows 
Kippenhahn diagrams for the baseline models post core He-burning. 
The C-burning features of both baseline models are similar; they
both ignite carbon convectively at the core and undergo three convective
C-burning flashes that recede outward in mass coordinate.

Post C-depletion, the photodisintegration of $^{20}$Ne drives
convective core Ne-burning. This burning phase lasts $\simeq$\,1.7
years for the solar model and $\simeq$\,0.33 years for the subsolar model.
After Ne-depletion, core O-burning begins at $T_{\rm{c}}$\,$\simeq$\,1.8$\times$10$^9$~K
and $\rho_{\rm{c}}$\,$\simeq$\,9.1$\times$10$^{6}$~g~cm$^{-3}$. The initial core O-burning 
episode is energetic enough to drive a large convection region that initially extends to
$\simeq$\,0.9\,\Msun.
At core O-depletion,
we find a composition of
$X_{\rm{c}}(^{32}\rm{S})$\,$\simeq$\,0.524,
$X_{\rm{c}}(^{34}\rm{S})$\,$\simeq$\,0.189, and
$X_{\rm{c}}(^{28}\rm{Si})$\,$\simeq$\,0.244 for the solar model.
Other isotopes show central mass fractions of
$X_{\rm{c}}$\,$\lesssim$\,10$^{-2}$.
The subsolar model has an O-depletion composition of
$X_{\rm{c}}(^{32}\rm{S})$\,$\simeq$\,0.522,
$X_{\rm{c}}(^{34}\rm{S})$\,$\simeq$\,0.175, and
$X_{\rm{c}}(^{28}\rm{Si})$\,$\simeq$\,0.236 
with other burning products
having negligible central mass fractions.
The central electron fraction at this point is
$Y_{\rm{e,c}}\simeq0.4936$
and
$Y_{\rm{e,c}}\simeq0.4942$
for the solar and subsolar models, respectively.
Our choice of stopping criterion does not signify the end of O-burning.

\begin{figure*}[!htb]
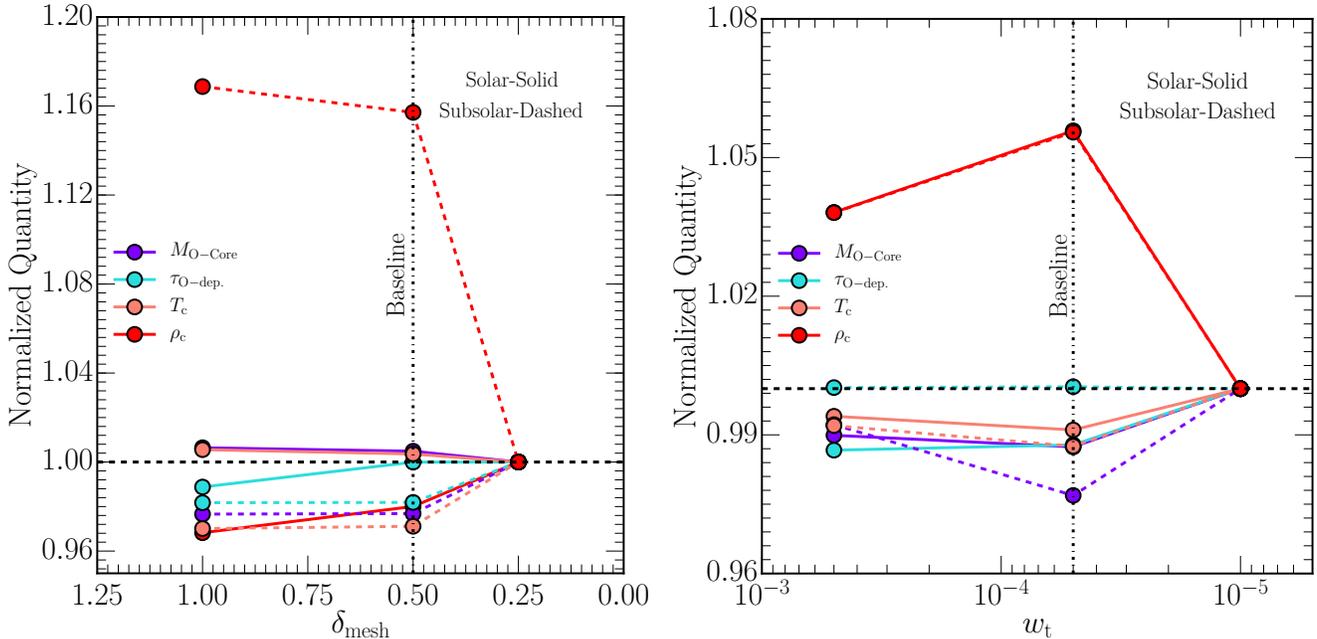

\begin{subfigure}
\centering{\includegraphics[width=\columnwidth]{{{conv_mesh_picked-cropped}}}}
\end{subfigure}
\begin{subfigure}
\centering{\includegraphics[width=\columnwidth]{{{conv_var_picked-cropped}}}}
\end{subfigure}
\caption{
Four normalized quantities at core O-depletion as a function of the 
mass (left) and temporal (right) resolution
controls $\delta_{\rm{mesh}}$ and $w_{\rm{t}}$:
mass of the oxygen core - $M_{\rm{O-Core}}$,
age - $\tau_{\rm{O-dep.}}$,
central temperature - $T_{\rm{c}}$, and
central density - $\rho_{\rm{c}}$.
All quantities are normalized to their values at 
$\delta_{\rm{mesh}}=0.25$
and $w_{\rm{t}}=1\times10^{-5}$.
A vertical black dash-dot line marks the resolution used for 
the Monte Carlo stellar models.
A black dashed horizontal line marks a value of 
unity, i.e. no variance with respect
to changes in mass or temporal resolution.
Solid lines correspond to
the solar models and dashed lines to the subsolar models.}
\label{fig:baseline_conv}
\end{figure*}

Lastly, we consider the impact of mass and temporal resolution on key
physical parameters relevant to this paper by evolving eight additional 
baseline models. Figure~\ref{fig:baseline_conv} shows the results for
$\delta_{\rm{mesh}}$\,=\,(1.0, 0.25) at our fixed baseline
temporal resolution of $w_{\rm{t}}$\,=\,5$\times$10$^{-5}$, and
$w_{\rm{t}}$\,=\,(5$\times$10$^{-4}$, 1$\times$10$^{-5})$ at our baseline mass
resolution of $\delta_{\rm{mesh}}$\,=\,0.5. Otherwise the solar and subsolar models 
use the same \emph{median} reaction rates and input physics as the baseline
models. For \mesh, the largest variation is $\simeq 13\,\%$
in the central density for the subsolar models. All other quantities have variations
$\lesssim 7\%$ at the highest mass resolution considered. For  $w_{\rm{t}}$,
the largest variation is $\lesssim$\,5\% in the central density, and
all other quantities have variations of $\lesssim$\,3\%.

\section{Monte Carlo Stellar Models}
\label{sec:mcstars}

We evolve two grids of Monte Carlo stellar models. The first grid
consists of 1000 Monte Carlo stellar models at solar metallicity. Each
model has a different set of sampled nuclear reactions; otherwise each model
has the same input physics as the baseline model.
We refer to this set of models as the ``solar grid''. The second set consists
of 1000 models at a metallicity of $Z$=0.0003, henceforth the ``subsolar grid''.
Each stellar model takes $\simeq$\,60 hours on 4 CPUs. The total 
computational expense is $\simeq$\,0.48 M CPU hours and generates 
$\simeq$\,1~TB of data.

Some properties of a stellar model may be more important
at different evolutionary phases. For example,
the time spent on the main-sequence is a direct consequence of
the $^{14}$N$(p,\gamma)^{15}$O reaction which modulates the rate at which
the CNO cycle may proceed \citep{imbriani_2004_aa}. At core He-depletion,
the central carbon mass fraction, temperature, or density affects whether
carbon ignites radiatively or convectively 
\citep{lamb_1976_aa,woosley_1986_aa, petermann_2017_aa}.
Such features are directly linked to key nuclear reaction rates. We thus consider different
properties of our stellar models at five evolutionary epochs: central H-, He-, C-,
Ne-, and O-depletion. The properties considered at each epoch are
commonly held to be significant for connecting presupernova stellar models to
observed transients, stellar yields for chemical evolution, or predicting SN properties
\citep[e.g.,][]{nomoto_2013_aa,couch_2015_aa,janka_2016_aa,cote_2017_aa}.

To determine the reaction rates that have the largest impact on
different properties of the stellar models at different evolutionary
phases, we use a Spearman Rank-Order Correlation (SROC) analysis.
A SROC is the Pearson correlation
coefficient between the rank values of two variables \citep{myers_2003_aa}.
The $N$ raw scores $A_i$ and $B_i$ are converted to ranks rg$A_i$ and rg$B_i$,
sorted in descending order according to magnitude, and the SROC is
\begin{equation}
r_{\rm{s}} = \frac{\textup{cov}(\textup{rg}A ,\textup{rg}B)}{\sigma_{\textup{rg}A}\sigma_{\textup{rg}B}}~,
\end{equation}
where $\textup{cov}(\textup{rg} A,\textup{rg} B)$ is the covariance matrix of the two variables
$A_i$ and $B_i$,
and $\sigma_{\textup{rg} A}$ and $\sigma_{\textup{rg} B}$ are the standard deviations of $A$
and $B$, respectively
A SROC of $r_{\rm{s}}$\,=\,+1 represents a perfectly
monotonically increasing relationship, $r_{\rm{s}}$\,=\,0, perfectly uncorrelated, and
$r_{\rm{s}}$\,=\,$-$1, monotonically decreasing.

\subsection{Hydrogen Depletion}
\label{sec:h_dep}

\begin{figure*}[!htb]
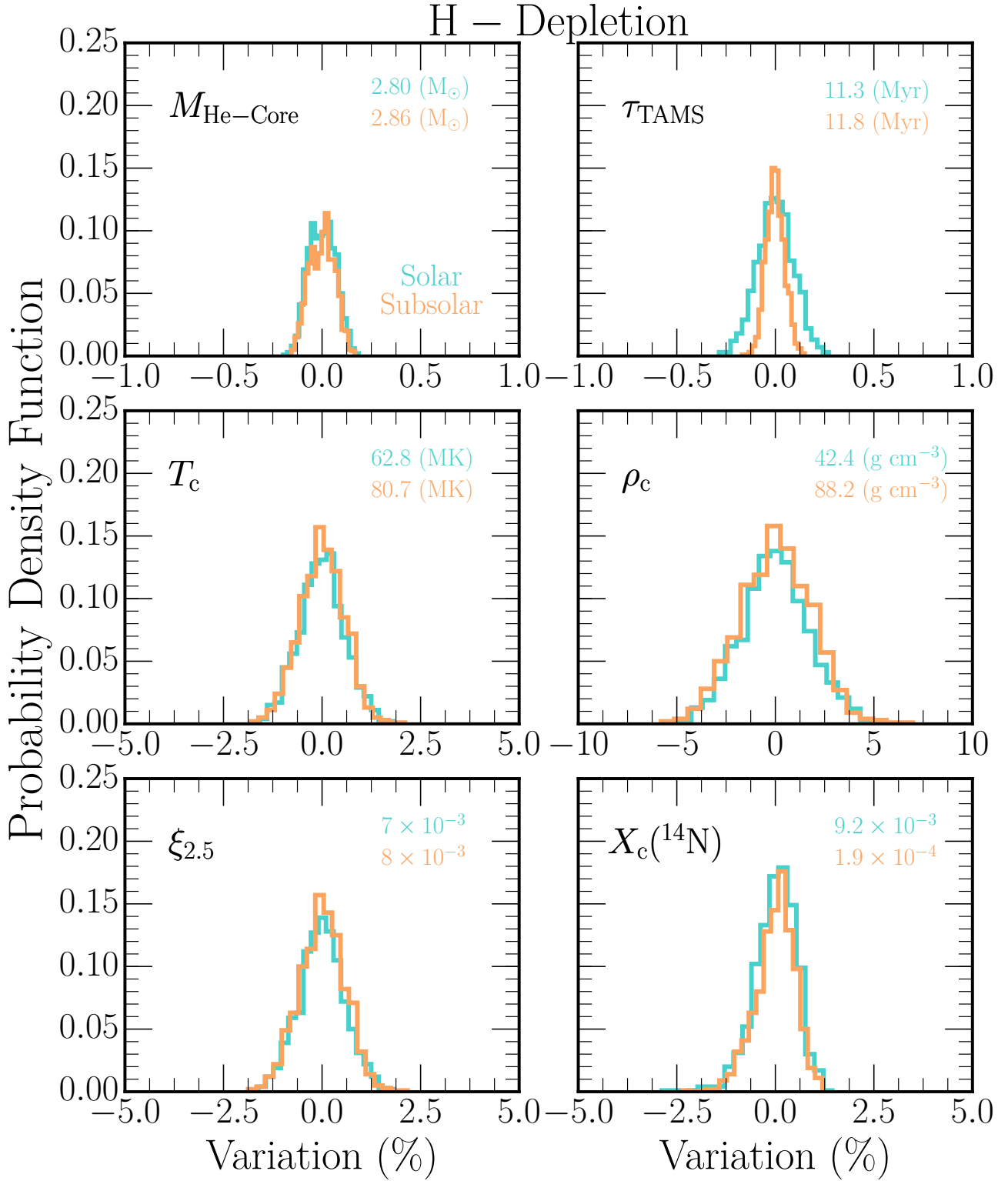

\centering{\includegraphics[width=2.0\columnwidth]{{{all_pdfs_h_dep_v5-cropped}}}}
\caption{
Probability density functions for six properties of the grid of Monte Carlo stellar models.
The x-axis represents the difference of a model value for a given property and the 
arithmetic mean of all values obtained for that property. This quantity is then normalized
to the mean of the distribution. This distribution is referred to as the 
``variation''. The blue histograms correspond to the solar 
models while the tan histograms denote the
subsolar models. The properties shown are
$M_{\rm{He-Core}}$ - the mass of the He core,
$\tau_{\rm{TAMS}}$ - the age at hydrogen depletion,
$T_{\rm{c}}$ - central temperature,
$\rho_{\rm{c}}$ - central density,
$\xi_{2.5}$ - compactness parameter measured at m\,=\,2.5\,\Msun,
and
$X_{\rm{c}}(^{14}\rm{N})$ - central $^{14}$N mass fraction.
All properties are measured at H-depletion, when X($^{1}$H) $\lesssim$ 10$^{-6}$.
Annotated are the arithmetic means of each property corresponding to a variation of zero.
}
\label{fig:all_pdfs_h_dep}
\end{figure*}

\begin{figure*}[!htb]
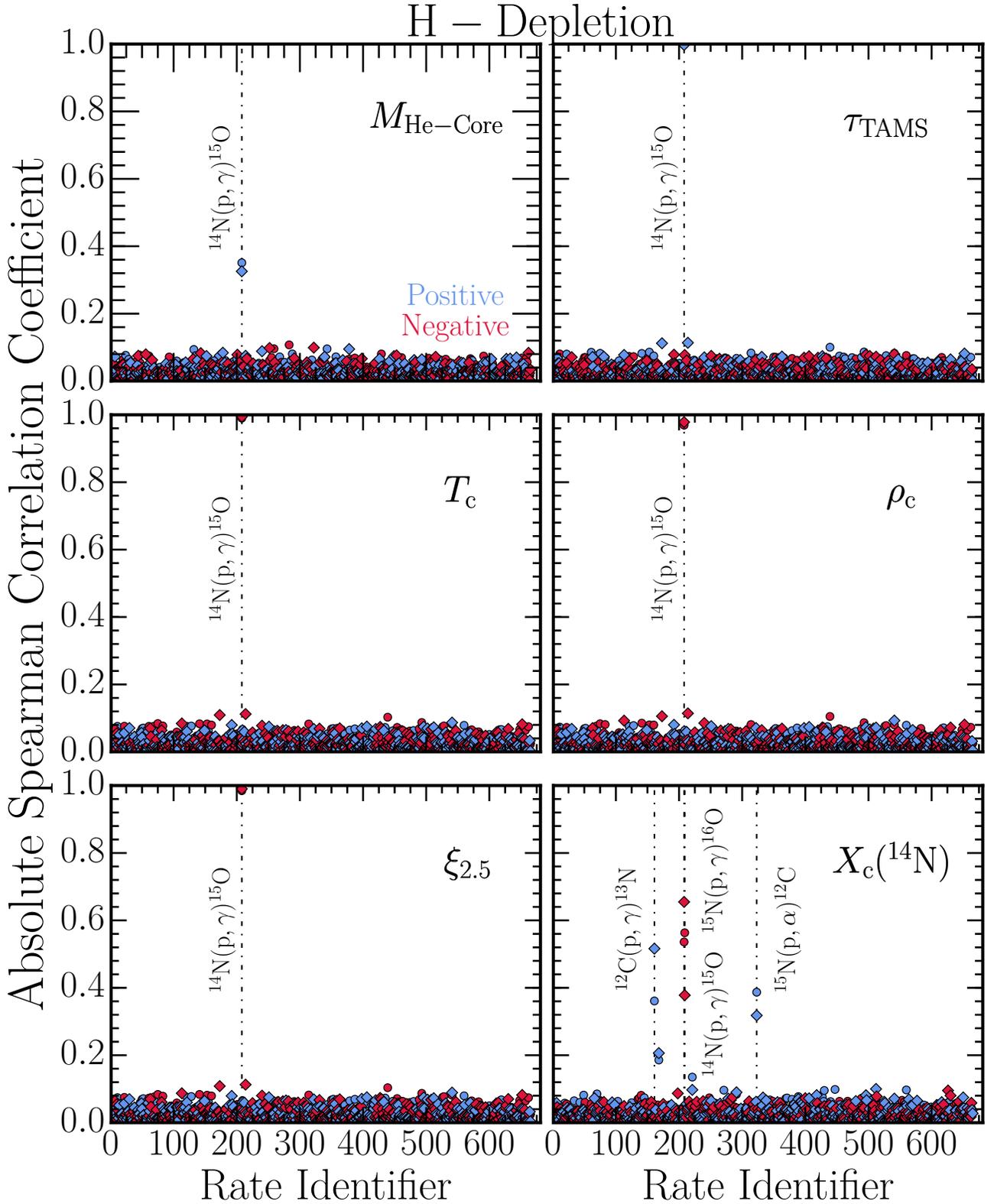
\centering{\includegraphics[width=2.0\columnwidth]{{{all_rhos_h_dep_v5-cropped}}}}
\caption{
The absolute Spearman Rank-Order Correlation Coefficients for the 665 independently
sampled thermonuclear reaction rates for the solar and subsolar grid of 1,000 Monte
Carlo stellar models. For a single nuclear reaction, the array of 1,000 Gaussian deviates
that were used to construct the sampled rate distributions is compared to six properties
of the stellar models to determine the correlation coefficient. The solar metallicity
coefficients are denoted by circles and subsolar metallicity by diamonds. A positive
correlation is denoted by a blue marker while a negative coefficient is represented by
a red marker. The x-axis corresponds to an arbitrary ``rate identifier'' used to track
the thermonuclear reaction rates sampled in this work. 
The quantities considered are
$M_{\rm{He-Core}}$ - the mass of the He core,
$\tau_{\rm{TAMS}}$ - the age,
$T_{\rm{c}}$ - the central temperature,
$\rho_{\rm{c}}$ - the central density,
$\xi_{2.5}$ - the compactness parameter,
and
$X_{\rm{c}}(^{14}$N) - the central neon-14 mass fraction.
All quantities are measured when $X_{\rm{c}}(^{1}\textup{H})$
$\lesssim 1\times10^{-6}$.
Key nuclear reactions with large SROC values are annotated.
}
\label{fig:all_rhos_h_dep}
\end{figure*}

We consider six properties at core H-depletion,
which we define as the time point when the central $^{1}$H mass fraction
drops below $\simeq$ 10$^{-6}$:
the mass of the He core $M_{\rm{He-Core}}$,
age  $\tau_{\rm{TAMS}}$,
central temperature $T_{\rm{c}}$,
central density $\rho_{\rm{c}}$,
compactness parameter, effectively the depth of the gravitational potential well
at the expected maximum mass of a neutron star, $\xi_{2.5}$=$M/R|_{m=2.5\,\Msun}$,
and central $^{14}$N mass fraction $X_{\rm{c}}(^{14}\rm{N})$.

\subsubsection{Probability Distribution Functions}
\label{s.pdf_h}

Figure~\ref{fig:all_pdfs_h_dep} shows the PDFs
of these six properties of the stellar models at this epoch.
The x-axis is the variation,
$
(X_i-\bar{X}) / \bar{X},
\label{eq:variation}
$
where $X_i$ is a value of a property for a single model and
$\bar{X}$ is the arithmetic mean of the distribution. The amplitude of
the histogram corresponds to the fraction of the 1000 models within a
given bin.  In this paper, the number of
bins is chosen according to the Rice Rule, $k=2n^{1/3}$, where $k$ is
the number of bins and $n$ is the number of samples
\citep[]{rice_rule}. While different bin widths can reveal
different features of the distribution, we find this choice of 
bins sufficient for the discussion of the histograms presented here.

\begin{figure}[!htb]
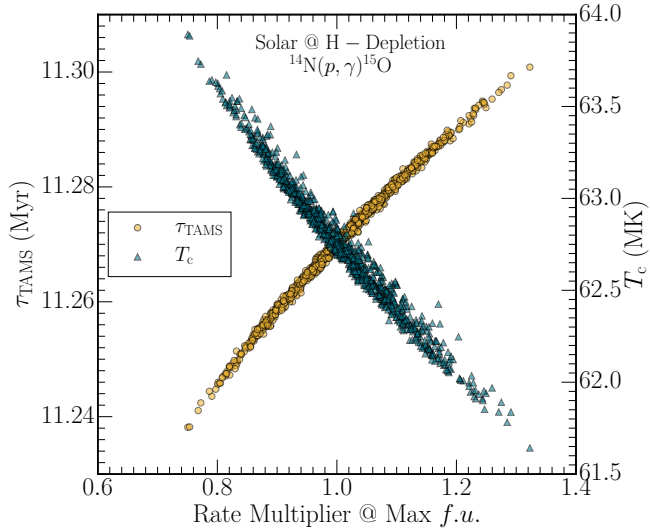

\centering{\includegraphics[width=1.0\columnwidth]
{{{n14pg_v_age_v_T_solar_v2-cropped}}}}
\caption{
The age and central temperature at H-depletion as
a function of the max rate multiplier applied to the
$^{14}$N($p,\gamma$)$^{15}$O reaction rate for the
grid of solar metallicity stellar models. The max $f.u.$
for $^{14}$N($p,\gamma$)$^{15}$O is $\simeq$ 1.1.
Best-fit lines to the two trends yield the slope
of the thermostat mechanism to be 
$\textup{d} \tau_{\rm{TAMS}} / \textup{d}T_{\rm{c}}$
$\simeq -0.03$ Myr/MK.}
\label{fig:n14pg_age_T_solar_h_dep}
\end{figure}

Throughout this paper we use the 95\% Confidence Interval (CI) limits.
These are defined, for each PDF, to be the limits corresponding to the unique
cumulative distribution function containing 95\% of the PDF.
This allows reporting the most likely ($\sim2\sigma$) values of a property
without the effects of outliers in the data. This definition is different
than a canonical CI derived from an assumed distribution function model of the data.

We define $M_{\rm{He-Core}}$ as the mass coordinate where $X(^1$H)$<$
0.01 and $X(^4$He)$>$0.1. The 95\% CI widths of
the $M_{\rm{He-Core}}$ PDFs span a narrow $\simeq$\,$\pm$\,0.1\% across the mean
of the distribution for both solar and subsolar models.  Both PDFs show well-defined
zero variation peaks of 2.80\,\Msun for the solar models and
2.86\,\Msun for the subsolar models.

The 95\% CI width of the $\tau_{\rm{TAMS}}$ PDF for the solar models,
$\simeq \pm$~0.2\%, is larger than the width of the PDF for the
subsolar models, $\simeq \pm$~0.1\%. We defer an explanation
of this difference until we discuss Figure \ref{fig:all_rhos_h_dep}.
The solar and subsolar PDFs are symmetric about
their zero variation values of 11.3\,Myr and 11.8\,Myr, respectively.

The $T_{\rm{c}}$ and $\rho_{\rm{c}}$ PDFs show the
solar models are slightly cooler and less dense than
the subsolar models, with zero variation values of $T_{\rm{c}}
\simeq$~(62.8 MK, 80.7 MK) and $\rho_{\rm{c}} \simeq$~(42.4 g
cm$^{-3}$, 88.2 g cm$^{-3}$), respectively. After H-depletion, the
solar models will proceed to burn He at a cooler core temperature but
more dense core. This trend is seen in Figure~\ref{fig:baseline_TRho}.
Note the subsolar models have larger $T_{\rm{c}}$ and yet {\it longer}
lifetimes $\tau_{\rm{TAMS}}$.  In addition, the 95\% CI width of the
$T_{\rm{c}}$ PDFs are $\simeq$\,1.2\%, and the 95\% CI width of the
$\rho_{\rm{c}}$ PDFs are $\simeq$\,4\%.

Traditionally $\xi_{2.5}$ is evaluated at core collapse.
Our motivation for measuring $\xi_{2.5}$ starting
at H-depletion is to assess the evolution of the variability 
in $\xi_{2.5}$; when do significant variations first seed
and how do the variation grow.
The 95 \% width of the $\xi_{2.5}$ PDF at H-depletion, $\simeq$ 1.2\%, 
is dependent upon the narrow $M_{\rm{He-Core}}$ PDF and the wider $\rho_{\rm{c}}$
PDF.  In addition, $\xi_{2.5}$ depends on the gradient of the density profile.  
The zero variation values of the solar and
subsolar grids show small differences at this epoch with $\xi_{2.5}
\simeq$~(7$\times$10$^{-3}$, 8$\times$10$^{-3}$), respectively.

Nitrogen is the dominant metal in the ashes of H-burning in massive
stars because the $^{14}$N($p,\gamma$)$^{15}$O rate is the smallest in
the CNO cycles \citep[e.g.,][]{iben_1966_aa}. This is reflected in the
$X_{\rm{c}}(^{14}\rm{N})$ PDFs by the zero variation values,
9.2$\times$10$^{-3}$ for the solar models and 1.9$\times$10$^{-4}$ for
the subsolar models, being approximately equal to the sum of the ZAMS
CNO mass fractions.  The 95\% CI width of the $X_{\rm{c}}(^{14}\rm{N})$ PDF,
$\simeq \pm$~1\%, is consistent with the spreads in the other
quantities measured.

\subsubsection{Spearman Correlation Coefficients}
\label{s.sroc_h}

Figure~\ref{fig:all_rhos_h_dep} shows the SROC coefficients for the solar
and subsolar grids. The coefficients for
the solar grid is shown by circles while the subsolar grid is shown by diamond
markers. A positive correlation coefficient is represented by a blue marker,
while a negative coefficient is denoted by a red marker.  For each property shown,
the rate identifier corresponding to the largest magnitude SROC
coefficients are marked by a vertical dashed line and label. 

The $^{14}$N($p,\gamma$)$^{15}$O rate has a large impact on all the
quantities we measure. For example, the $^{14}$N($p,\gamma$)$^{15}$O
rate has the largest SROC coefficient for $\tau_{\rm{TAMS}}$,
with $r_{\rm{s}}$\,$\simeq$+0.99 for the solar and subsolar models.
Coefficients of the remaining 664 reactions are significantly smaller,
$\mathcal{O}(10^{-2})$.  This suggest that $\tau_{\rm{TAMS}}$ is a
directly dependent on the $^{14}$N($p,\gamma$)$^{15}$O rate, with a
larger rate {\it increasing} the lifetime to core H-depletion
\citep[e.g.,][]{imbriani_2004_aa,weiss_2005_aa,herwig_2006_ab}.

Increasing a reaction rate usually increases the nuclear energy generation
rate, which deposits its energy into thermal energy. The core
temperature rises.  Via the equation of state, the pressure increases,
which causes the stellar core to expand.  This expansion decreases
$T_{\rm{c}}$ and $\rho_{\rm{c}}$, and thus causes nuclear burning to proceed
at a slower rate. The net result of increasing an energetically important
reaction rate is a longer burning lifetime and a decreased 
$T_{\rm{c}}$ and $\rho_{\rm{c}}$. This is the well-known thermostat 
mechanism \citep[e.g.,][]{hansen_2004_aa, iliadis_2007_aa}.

Figure \ref{fig:n14pg_age_T_solar_h_dep} shows the age and $T_{\rm{c}}$ at
H-depletion for the solar models as a function of the rate multiplier
applied at max $f.u.$ for the $^{14}$N($p,\gamma$)$^{15}$O reaction.
Least-square fits to the linear trends yield the slope
of the thermostat mechanism:
$\textup{d} \tau_{\rm{TAMS}} / \textup{d}T_{\rm{c}}$, $\simeq$\,$-0.03$\,Myr/MK.
This correlation is confirmed by the large and negative SROC
coefficients between the $^{14}$N($p,\gamma$)$^{15}$O rate
and $\xi_{2.5}$, $T_{\rm{c}}$, and $\rho_{\rm{c}}$.  The thermostat
mechanism also causes the slightly larger zero variation of
$M_{\rm{He-Core}}$ for the subsolar models relative to the solar models
in Figure~\ref{fig:all_pdfs_h_dep}.

\subsubsection{Impact of the Measurement Point}
\label{s.meas_h}

To assess the impact of the choice of the measurement point,
we repeat our SROC analysis during core H-burning at the point
$X_{\rm{c}}(^{1}\rm{H})$\,$\simeq$\,$X_{\rm{c}}(^{4}\rm{He})$.
We compare the magnitude of the SROC values for
$\tau_{\rm{TAMS}}$,
$T_{\rm{c}}$,
$\rho_{\rm{c}}$, and
$\xi_{2.5}$ for both the solar and subsolar models.

Qualitatively, $^{14}$N($p,\gamma$)$^{15}$O still
drives the variation in the age with a positive correlation, and
the variations in
$T_{\rm{c}}$,
$\rho_{\rm{c}}$, and
$\xi_{2.5}$ with negative correlations. The difference of the SROC values
between the two epochs agree to $\lesssim 0.01$ for $T_{\rm{c}}$,
$\rho_{\rm{c}}$, and
$\xi_{2.5}$ and $\lesssim 0.2$ for $\tau_{\rm{TAMS}}$.
This re-evaluation suggests the PDFs vary slightly based on the 
chosen measurement point and identifying the key reactions from 
the SROC analysis is an invariant.

\subsection{Helium Depletion}
\label{sec:he_dep}

\begin{figure*}[!htb]
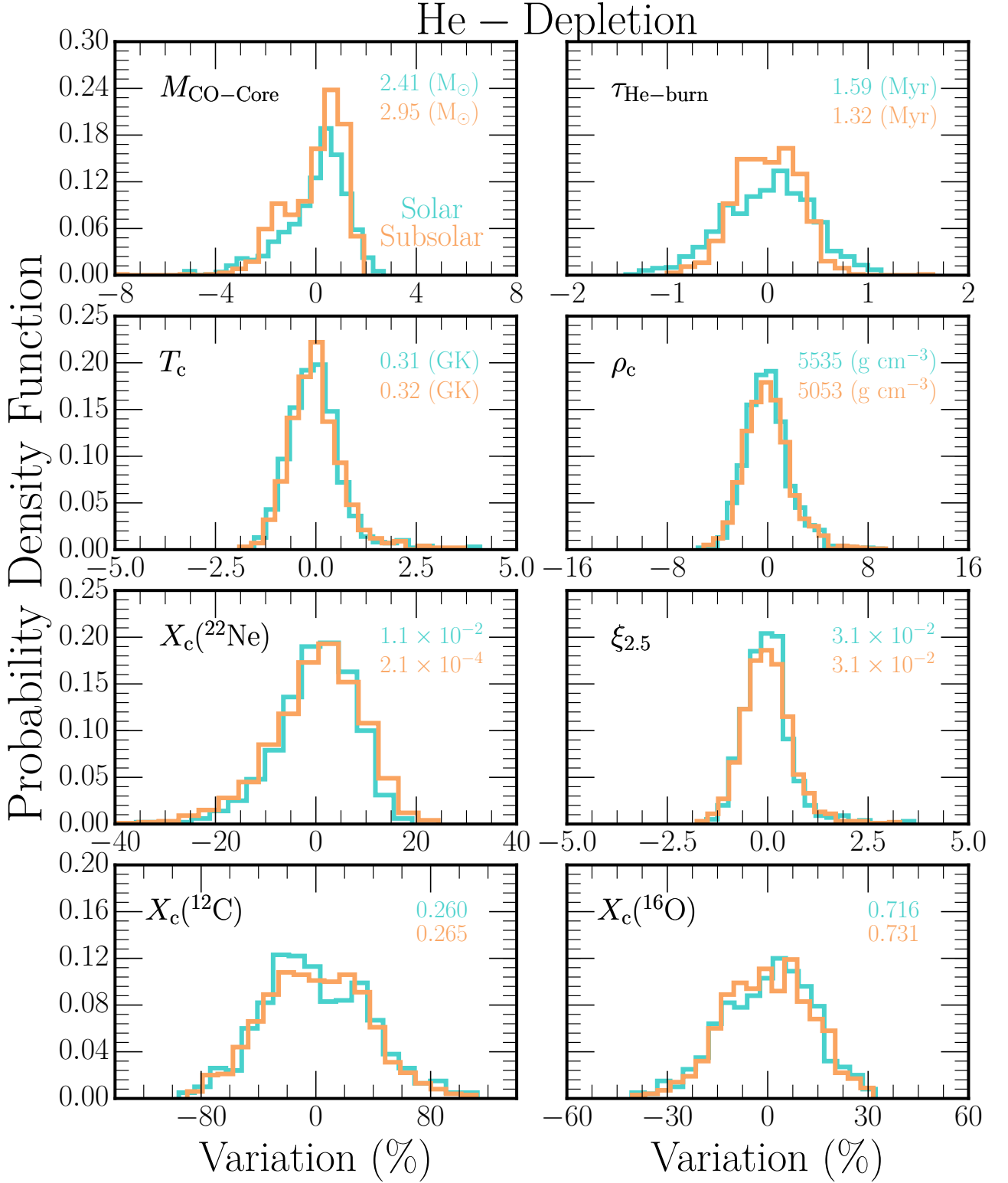

\centering{\includegraphics[width=2.0\columnwidth]{{{all_pdfs_he_dep_v4-cropped}}}}
\caption{
Same as in Figure~\ref{fig:all_pdfs_h_dep} except we consider
$M_{\rm{CO-Core}}$ - the mass of the CO core,
$\tau_{\rm{He-burn}}$ - the elapsed time between H-depletion and He-depletion,
$T_{\rm{c}}$ - the central temperature,
$\rho_{\rm{c}}$ - the central density,
$X_{\rm{c}}(^{22}$Ne) - the central neon-22 mass fraction,
$\xi_{2.5}$ - the compactness parameter,
$X_{\rm{c}}(^{12}$C) - the central carbon-12 mass fraction,
and
$X_{\rm{c}}(^{16}$O) - the central oxygen-16 mass fraction
all measured at He-depletion.
}
\label{fig:all_pdfs_he_dep}
\end{figure*}

\begin{figure*}[!htb]
\centering{\includegraphics[width=2.0\columnwidth]{{{all_rhos_he_dep_v4-cropped}}}}
\caption{
Same as in Figure~\ref{fig:all_rhos_h_dep}. The quantities considered are
$M_{\rm{CO-Core}}$ - the mass of the CO core,
$\tau_{\rm{He-burn}}$ - the elapsed time between H-depletion and He-depletion,
$T_{\rm{c}}$ - the central temperature,
$\rho_{\rm{c}}$ - the central density,
$X_{\rm{c}}(^{22}$Ne) - the central neon-22 mass fraction,
$\xi_{2.5}$ - the compactness parameter,
$X_{\rm{c}}(^{12}$C) - the central carbon-12 mass fraction,
and
$X_{\rm{c}}(^{16}$O) - the central oxygen-16 mass fraction.
All quantities used here were measured at He-depletion.
}
\label{fig:all_rhos_he_dep}
\end{figure*}

We measure the integrated impact of the uncertainties in the reaction rates
at the point when the central helium mass fraction X($^4$He) $\lesssim 10^{-6}$.

\subsubsection{Probability Distribution Functions}
\label{s.pdf_he}

Figure~\ref{fig:all_pdfs_he_dep} shows the PDFs of eight properties
from the stellar models at this epoch:
mass of the CO core $M_{\rm{CO-Core}}$,
the elapsed time between H-depletion and He-depletion $\tau_{\rm{He-burn}}$,
central temperature $T_{\rm{c}}$,
central density $\rho_{\rm{c}}$,
central $^{22}$Ne mass fraction $X_{\rm{c}}(^{22}$Ne),
compactness parameter $\xi_{2.5}$,
central $^{12}$C mass fraction $X_{\rm{c}}(^{12}$C), and
central $^{16}$O mass fraction $X_{\rm{c}}(^{16}$O).

The 95\% CI width of the $M_{\rm{CO-Core}}$ PDF spans $\simeq$\,$\pm$\,2\%
for the solar and subsolar grids. Both PDFs show a well-defined peak of 
2.41\,\Msun for the solar models and 2.95\,\Msun for the subsolar models 
and an extended tail for negative variations.  That is, changes in the
reaction rates are more likely to produce smaller C cores than more
massive C cores.  This asymmetry accounts for the PDFs not being
centered at zero variation.

The solar and subsolar grid PDFs for $\tau_{\rm{He-burn}}$ have a 95\%
CI spread of $\simeq \pm~1\%$, suggesting rate
uncertainties have a smaller impact on $\tau_{\rm{He-burn}}$.  The
solar PDF is slightly wider the subsolar PDF, and both PDFs are
symmetric about their respective arithmetic means.

The $T_{\rm{c}}$ and $\rho_{\rm{c}}$ PDFs show 95\% CI widths
of $\simeq \pm~1.5\%$ and $\simeq \pm~3.5\%$, respectively,
for both solar and subsolar models. Both PDFs are centrally
peaked with $\lesssim$\,1\% differences between the arithmetic means of the
solar and subsolar models. Both PDFs exhibit long tails
in the positive variation direction, indicating some combinations
of the reaction rates produce cores that are $\simeq$5\% hotter than
the mean and $\simeq$10\% denser than the mean.

The solar and subsolar grid PDFs for $X_{\rm{c}}$($^{22}$Ne) PDF are nearly the same.
However, the arithmetic mean of the two PDFs differ by a factor of $\simeq$~50.
The reason for this difference is that most of a ZAMS star's initial
metallicity $Z$ comes from the CNO and $^{56}$Fe nuclei inherited from
its ambient interstellar medium. The slowest step in the hydrogen burning CNO cycle is
$^{14}$N($p,\gamma$)$^{15}$O, which causes all the CNO catalysts
to pile up at $^{14}$N at core H-depletion. During He-burning the sequence
$^{14}$N($\alpha$,$\gamma$)$^{18}$F($\beta^{+},\nu_e$)$^{18}$O($\alpha$,$\gamma$)$^{22}$Ne
converts all of the $^{14}$N into the neutron-rich isotope $^{22}$Ne.
Thus, $X_{\rm{c}}$($^{22}$Ne) at core He-depletion is linearly
dependent on the initial CNO abundances. The subsolar models
have $\simeq$\,50 times less initial CNO than the solar models,
accounting for the difference in the arithmetic means.

\begin{figure}[!htb]
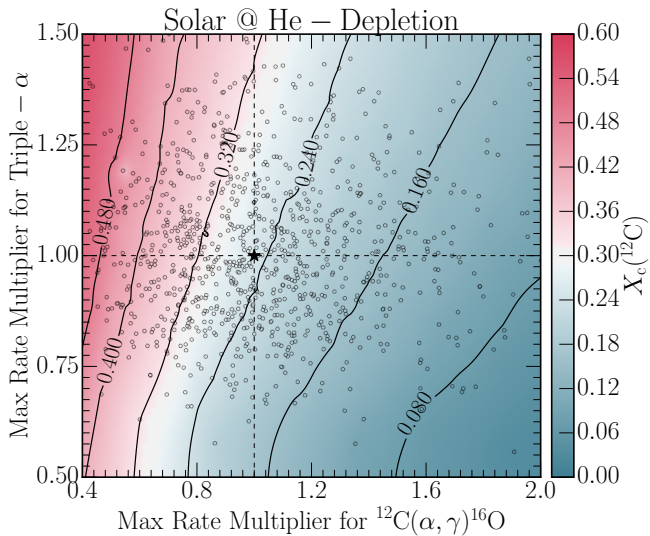

\centering{\includegraphics[width=\columnwidth]
{{{c12_v_c12ag_v_3a_he_dep_solar_v3-cropped}}}}
\caption{
Central carbon mass fraction at helium depletion for solar models 
as a function of the rate multiplier at max $f.u.$ applied to the
$^{12}$C($\alpha,\gamma$)$^{16}$O and triple-$\alpha$ rates.
The heatmap uses bi-linear interpolation and extrapolation
of the models, which are shown by gray circles.
Contour lines of constant $X_{\rm{c}}$($^{12}$C)  shown by
solid black lines. Also shown by a dashed line is the value of a rate multiplier
of unity for both reactions. Lastly, the black star denotes the value of
$X_{\rm{c}}(^{12}\rm{C})$ found for the \emph{median} reaction rates.
Compare with Figure~20 of \citet{west_2013_aa}.}
\label{fig:c12_v_c12ag_v_3a_he_dep_solar}
\end{figure}

The solar and subsolar PDFs for $\xi_{2.5}$
are similar in peak amplitude, 95\% CI width ($\simeq$\ 1.2\%),
symmetry about zero variation, and mean arithmetic value.  That is,
rate uncertainties have little impact on differentiating
between solar and subsolar metallicities. Similar to the
$T_{\rm{c}}$ and $\rho_{\rm{c}}$ PDFs, there are outlier models
whose reaction rate combinations produce larger $\xi_{2.5}$.

The largest variations occur in the $X_{\rm{c}}(^{12}$C) and
$X_{\rm{c}}(^{16}$O) PDFs with 95\% CI widths of $\simeq$\,$\pm$\,70\% and
$\simeq$\,$\pm$\,25\%, respectively.  The common driver for these
variations are the triple-$\alpha$, $^{12}$C($\alpha,\gamma$)$^{16}$O,
and $^{16}$O($\alpha,\gamma$)$^{20}$Ne rates, whose roles we
discuss below.

\subsubsection{Spearman Correlation Coefficients}
\label{s.sroc_he}

Figure~\ref{fig:all_rhos_he_dep} shows the SROCs for the 665
independently sampled thermonuclear reaction rates against the eight
quantities considered in Figure~\ref{fig:all_pdfs_he_dep}.
The $M_{\rm{CO-Core}}$ is chiefly set by the
$^{12}$C($\alpha,\gamma$)$^{16}$O rate with 
$r_{\rm{s}}$\,$\simeq$+0.8 for both metallicity grids.
Larger $^{12}$C($\alpha,\gamma$)$^{16}$O rates build larger CO core masses.
The triple-$\alpha$ rate plays a smaller role
with $r_{\rm{s}}$\,$\simeq$\,$-0.17$ for both metallicity grids.
Similarly, $\tau_{\rm{He-burn}}$ is
primarily set by the $^{12}$C($\alpha,\gamma$)$^{16}$O rate
with coefficients of $r_{\rm{s}}=(+0.92,+0.94)$, respectively.
The triple-$\alpha$ rate plays a less
significant role with $r_{\rm{s}}$\,$\simeq$\,$ -0.25$.

In contrast, $T_{\rm{c}}$ and $\rho_{\rm{c}}$ are chiefly affected by
the uncertainties in the triple-$\alpha$ rate with
$r_{\rm{s}}$\,$\simeq$\,$-0.8$ and $r_{\rm{s}}$\,$\simeq$\,$-0.7$, respectively. These
large negative SROCs mean the thermostat mechanism,discussed
for H-burning, namely larger energy producing reaction
rates yield cooler and less dense cores, operates during He burning.
The $^{12}$C($\alpha,\gamma$)$^{16}$O and
$^{16}$O($\alpha,\gamma$)$^{20}$Ne rates play smaller roles
with $r_{\rm{s}}$\,$\le$\,+0.4. Note that the positive
correlation means larger $^{12}$C($\alpha,\gamma$)$^{16}$O rates produce 
hotter cores, in juxtaposition to the triple-$\alpha$ rate.
This is because a larger $^{12}$C($\alpha,\gamma$)$^{16}$O converts
more carbon into oxygen, so the core burns hotter at any given triple-$\alpha$ rate 
(which dominates the energy generation) to satisfy the 
luminosity demanded by the surface of the stellar model.
Outliers with positive variations in the $T_{\rm{c}}$ and $\rho_{\rm{c}}$ PDFs  of
Figure~\ref{fig:all_pdfs_he_dep} are caused by combinations of the
$^{12}$C($\alpha,\gamma$)$^{16}$O and triple-$\alpha$ reactions. For a small
triple-$\alpha$ rate, the model will be hotter and more dense. When this is coupled
with a large $^{12}$C($\alpha,\gamma$)$^{16}$O rate, the stellar models at He-depletion
have a hotter and denser core with $T_{\rm{c}}$ increased by $\simeq$\,+5\% and
and  $\rho_{\rm{c}}$ increased by $\simeq$\,+10\%.

The mass fraction of the neutron-rich $^{22}$Ne isotope, is set by the competition between
the triple-$\alpha$ and $^{22}$Ne($\alpha$,$\gamma$)$^{26}$Mg rates.
The triple-$\alpha$ rate sets $T_{\rm{c}}$ and $\rho_{\rm{c}}$,
with a larger rate giving cooler and denser cores that favor the production
of $^{22}$Ne by the sequence
$^{14}$N($\alpha$,$\gamma$)$^{18}$F($\beta^{+},\nu_e$)$^{18}$O($\alpha$,$\gamma$)$^{22}$Ne.
This is the origin of the positive SROC coefficient for the triple-$\alpha$ rate
in the solar and subsolar grids.
On the other hand, $^{22}$Ne($\alpha$,$\gamma$)$^{26}$Mg destroys $^{22}$Ne,
storing the neutron excess in $^{26}$Mg.
This accounts for the negative SROC coefficient of $^{22}$Ne($\alpha$,$\gamma$)$^{26}$Mg
for both metallicity grids.

$\xi_{2.5}$ is chiefly set by the triple-$\alpha$ rate with
$r_{\rm{s}}$\,=\,$-$0.83 and $r_{\rm{s}}$\,=\,$-$0.74 for the solar
and subsolar models respectively. A larger triple-$\alpha$ rate
produces a smaller $\xi_{2.5}$, due to the decrease in overall
density of the stellar core. For the subsolar models the
$^{16}$O($\alpha,\gamma$)$^{20}$Ne rate plays a smaller role
($r_{\rm{s}}$\,=\,$-0.34$), but also decreases $\xi_{2.5}$ as the
rate becomes larger.  The solar and subsolar PDFs for $\xi_{2.5}$
shows outliers with variations up to $\simeq$ 5\%.  These outliers
form from same combination of reaction rates that produce denser
stellar models. Namely, models with high $\xi_{2.5}$ have either a
depressed triple-$\alpha$ rate, an enhanced 
$^{12}$C($\alpha,\gamma$)$^{16}$O, or both.

During quiescent He-burning the
3$\alpha$-process and the $^{12}$C($\alpha,\gamma$)$^{16}$O reaction
burn with high efficiency through pronounced resonance mechanisms
\citep[e.g.,][]{deboer_2017_aa}. In contrast, the
$^{16}$O($\alpha,\gamma$)$^{20}$Ne reaction lacks any such resonance
enhancement in the stellar energy range making its rate comparatively
much lower.  This essentially prohibits significant He-burning
beyond $^{16}$O and maintains the $^{12}$C/$^{16}$O balance we observe today.

The $^{12}$C($\alpha,\gamma$)$^{16}$O rate sets
$X_{\rm{c}}(^{12}$C) and $X_{\rm{c}}(^{16}$O) for both solar and
subsolar models with $r_{\rm{s}}$\,$\simeq$\,$-0.95$ and
$r_{\rm{s}}$\,$\simeq$\,+0.95, respectively.  A larger
$^{12}$C($\alpha,\gamma$)$^{16}$O rate destroys more C and
produces more O.  The triple-$\alpha$ rate plays a smaller
role in setting $X_{\rm{c}}(^{12}$C) and $X_{\rm{c}}(^{16}$O) with
$r_{\rm{s}}$\,$\simeq$\,+0.29 and $r_{\rm{s}}$\,$\simeq$\,$-0.28$, respectively.
A larger triple-$\alpha$ rate produces more C and less O.
These results suggest $X_{\rm{c}}(^{12}$C) and $X_{\rm{c}}(^{16}$O)
are determined primarily by the uncertainties in these two
reaction rates.

\subsubsection{Triple-$\alpha$ and $^{12}$C($\alpha,\gamma$)$^{16}$O}
\label{s.3alpha}

Figure~\ref{fig:c12_v_c12ag_v_3a_he_dep_solar} shows $X_{\rm{c}}$($^{12}$C)
at He-depletion for the solar models as a function of the rate
multiplier at max $f.u.$ (over core He-burning temperatures) applied to the
$^{12}$C($\alpha,\gamma$)$^{16}$O and triple-$\alpha$ rates 
(see Figure~\ref{fig:fu_factors}).
A $^{12}$C($\alpha,\gamma$)$^{16}$O rate that is small relative to
median value and a triple-$\alpha$ rate that is large
relative to its median value produces a large
$X_{\rm{c}}$($^{12}$C). Conversely, a high $^{12}$C($\alpha,\gamma$)$^{16}$O
rate and a small triple-$\alpha$ rate produces a small
$X_{\rm{c}}$($^{12}$C).  When both rates are at the median value of their
respective PDFs, unity rate multipliers in
Figure~\ref{fig:c12_v_c12ag_v_3a_he_dep_solar},
$X_{\rm{c}}$($^{12}$C)\,$\simeq$\,0.26 (see Figure \ref{fig:all_pdfs_he_dep}).
The trend is commensurate with
\citet[][their Figure~20]{west_2013_aa}.

\subsubsection{Impact of the Measurement Point}
\label{s.meas_he}

Core He-burning is initiated by the triple-$\alpha$ reaction
releasing $\approx$\,7.27\, MeV of energy. At early times,
nuclear energy generation in the core is governed by this reaction rate.
The emergence of fresh $^{12}$C as a product of the triple-$\alpha$ reaction
allows $^{12}$C($\alpha,\gamma$)$^{16}$O
to convert the $^{12}$C ashes into $^{16}$O in a race between
the two reactions to consume the He fuel \citep[e.g.,][]{deboer_2017_aa}.
The $^{12}$C/$^{16}$O ratio is determined by these two reaction rates.

Due to this evolution, we re-evaluate our SROC coefficients midway
through the core He-burning process, when $X_{\rm{c}}(^{4}\rm{He})$\,$\simeq$\,0.5.
The structural properties $-$ $T_{\rm{c}}$, $\rho_{\rm{c}}$ and
$\xi_{2.5}$ $-$ agree qualitatively when comparing the midway and
depletion points of the solar models.  A midway
measurement point yields $\simeq$\,15\% stronger correlations.
The triple-$\alpha$ rate still drives the variations with
a negative SROC.  For $X_{\rm{c}}(^{12}$C) and
$X_{\rm{c}}(^{16}$O) the midway and He-depletion measurement points
for the solar models differ by $\Delta \left | r_{\rm{s}} \right |
\lesssim 2\%$ in the SROC values.

When measuring midway through the core He-burning process,
variations in $\tau_{\rm{He-burn}}$ for the solar models become mainly
driven by the $^{14}$N($p,\gamma$)$^{15}$O rate with a positive SROC
coefficient. An increase in this rate causes the stellar core to proceed
through core H-burning at lower $T_{\rm{c}}$.  When measuring
$\tau_{\rm{He-burn}}$ midway through He-burning, we find the
$^{14}$N($p,\gamma$)$^{15}$O rate also yields a negative SROC
coefficient for $T_{\rm{c}}$. Models with lower $T_{\rm{c}}$ proceed through
He-burning at a slower rate, hence increasing the helium burning lifetime $\tau_{\rm{He-burn}}$.

\begin{figure*}[!htb]
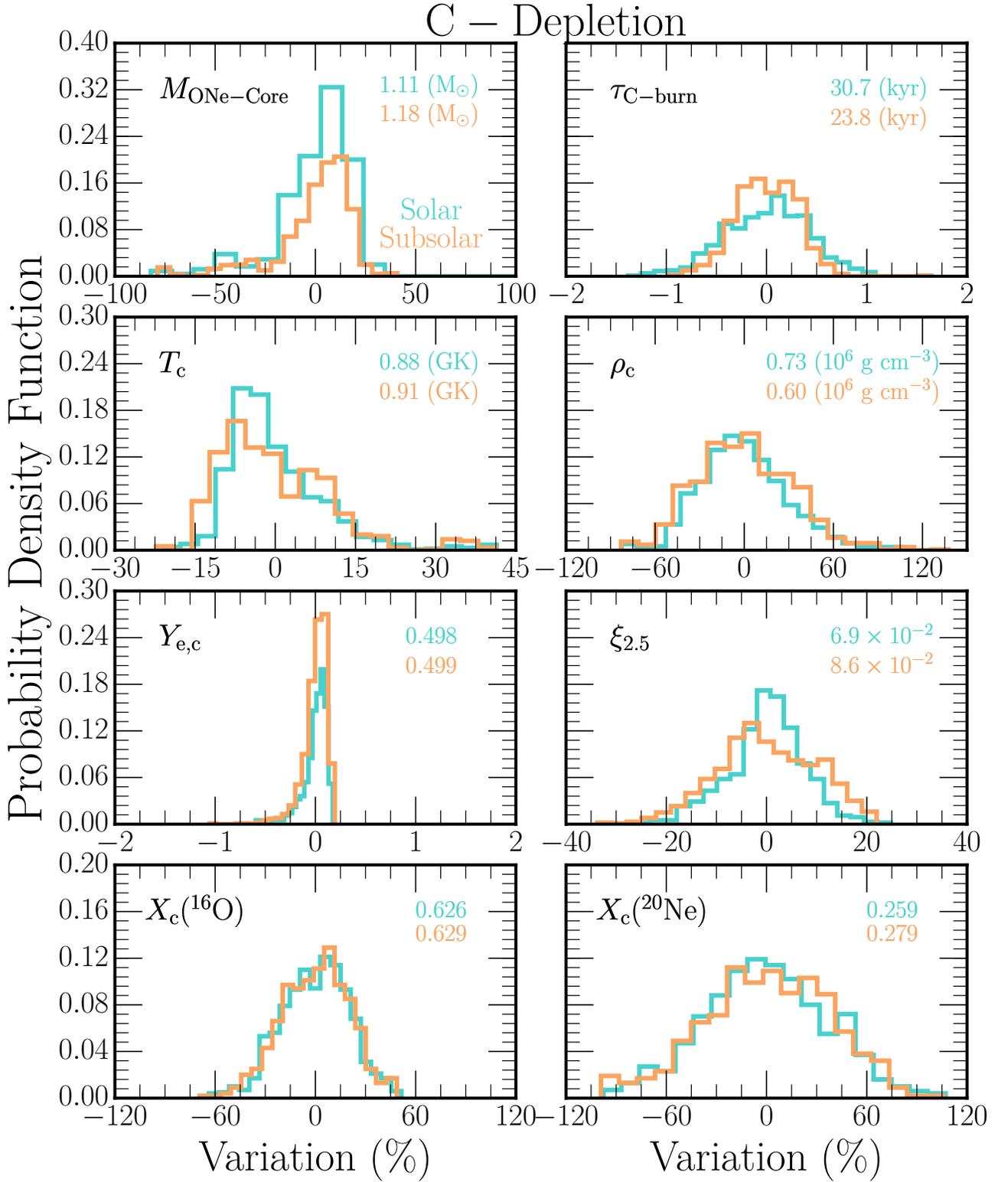

\centering{\includegraphics[width=2.0\columnwidth]{{{all_pdfs_c_dep_v4-cropped}}}}
\caption{
Same as in Figure~\ref{fig:all_pdfs_he_dep} except we consider
$M_{\rm{ONe-Core}}$ - the mass of the ONe core,
$\tau_{\rm{C-burn}}$ - the elapsed time between He-depletion to C-depletion,
$T_{\rm{c}}$ - the central temperature,
$\rho_{\rm{c}}$ - the central density,
$Y_{\rm{e,c}}$ - the central electron fraction,
$\xi_{2.5}$ - the compactness parameter,
$X_{\rm{c}}(^{16}$O) - the central oxygen-16 mass fraction,
and
$X_{\rm{c}}(^{20}$Ne) - the central neon-20 mass fraction.
All quantities are measured at C-depletion.
}
\label{fig:all_pdfs_c_dep}
\end{figure*}

\begin{figure*}[!htb]
\centering{\includegraphics[width=2.0\columnwidth]{{{all_rhos_c_dep_v5-cropped}}}}
\caption{
Same as in Figure~\ref{fig:all_pdfs_he_dep} except we consider
$M_{\rm{ONe-Core}}$ - the mass of the ONe core,
$\tau_{\rm{C-burn}}$ - the elapsed time between He-depletion and C-depletion,
$T_{\rm{c}}$ - the central temperature,
$\rho_{\rm{c}}$ - the central density,
$Y_{\rm{e,c}}$ - the central electron fraction,
$\xi_{2.5}$ - the compactness parameter,
$X_{\rm{c}}(^{16}$O) - the central oxygen-16 mass fraction,
and
$X_{\rm{c}}(^{20}$Ne) - the central neon-20 mass fraction.
All quantities are measured at C-depletion.
}
\label{fig:all_rhos_c_dep}
\end{figure*}

\subsection{Carbon Depletion}

Next, we measure the integrated impact of the reaction rate uncertainties at the point
when the central carbon mass fraction $X_{\rm{c}}(^{12}\textup{C}) \lesssim 1\times10^{-6}$.

\subsubsection{Probability Distribution Functions}
\label{s.pdf_c}

Figure~\ref{fig:all_pdfs_c_dep} shows the PDFs of eight properties of the 15\,\Msun
models at C-depletion:
mass of the ONe core $M_{\rm{ONe-Core}}$,
the elapsed time between He-depletion and C-depletion $\tau_{\rm{C-burn}}$,
central temperature $T_{\rm{c}}$,
central density $\rho_{\rm{c}}$,
central electron fraction $Y_{\rm{e,c}}$,
compactness parameter $\xi_{2.5}$,
central $^{16}$O mass fraction $X_{\rm{c}}(^{16}$O), and
central $^{20}$Ne mass fraction $X_{\rm{c}}(^{20}$Ne).

The $M_{\rm{ONe-Core}}$ distribution has 95\% CI variation limits of
$\simeq$\,+23\% and $\simeq$\,$-$50\% for the solar and subsolar models.
This is wider than the spread in the He core mass at H-depletion ($\simeq \pm~0.1\%$)
or the CO core mass at He-depletion ($\simeq \pm~3\%$).  We defer
explanation to Section \ref{s.meas_c} In
addition, the solar model PDF has a larger peak amplitude compared to
the subsolar model PDF.

In contrast, the 95\% CI spread of the $\tau_{\rm{C-burn}}$ distribution shows about the
same narrow width of $\simeq \pm$~1\% as $\tau_{\rm{TAMS}}$ and
$\tau_{\rm{He-burn}}$.  This is chiefly due to the CO core mass to be
burned laying within a relative narrow range ($\simeq$\,$\pm$\,3\%, see Figure
\ref{fig:all_pdfs_he_dep}). The solar model PDF has a zero
variance of $\tau_{\rm{C-burn}} \simeq$~30.7~kyr, while the subsolar model
PDF has a zero variance of $\tau_{\rm{C-burn}} \simeq$ 23.8~kyr.
This reflects the subsolar models undergoing hotter, less
dense core C-burning (see Figure \ref{fig:baseline_TRho}).

Carbon burning and the later stages of evolution in massive stars
have large core luminosities whose energy is carried away
predominantly by free-steaming neutrinos. These burning stages are thus
characterized by short evolutionary time scales.  When
thermal neutrinos instead of photons dominate the energy loss budget, carbon and
heavier fuels burn at a temperature chiefly set by the balanced power
condition $\left < \epsilon_{\rm nuc} \right > \simeq \left < \epsilon_{\nu} \right >$.
For core C-burning this gives $T_{\rm{c}}$ $\simeq$~0.9 GK and,
assuming a $T^3/\rho$ scaling,
$\rho_{\rm{c}}$~$\simeq$~6$\times$10$^{6}$ g cm$^{-3}$.  This is
commensurate with the zero variation values annotated in Figure
\ref{fig:all_pdfs_c_dep}.  The $T_{\rm{c}}$ and $\rho_{\rm{c}}$
distributions show 95\% CI widths of $\simeq$\,$\pm$\,15\% and $\simeq$\,$\pm$\,60\%
for the solar and subsolar models, respectively. This is
wider than the 95\% CI spreads of the $T_{\rm{c}}$ and $\rho_{\rm{c}}$
distributions at H-depletion and He-depletion.

The $Y_{\rm{e,c}}$ distributions show strong peaks at $Y_{\rm{e,c}} \simeq$~0.499
and 95\% CI spreads of $\lesssim 1\%$ for the solar and subsolar models. This is
commensurate with significant neutronization not occurring during quiescent core C-burning,
and shows $Y_{\rm{e,c}}$ is not strongly affected by the uncertainties in the reaction rates.

C-depletion marks the first occurrence of significant variation in
$\xi_{2.5}$. The solar and subsolar distributions show
95\% CI widths of $\simeq$\,$\pm$\,16\%. The mean value
of $\xi_{2.5} \simeq$~6.9$\times$10$^{-2}$
for the solar models is smaller than the mean value of $\xi_{2.5}
\simeq$~8.6$\times$10$^{-2}$ for the subsolar models.  This is due to
the smaller $\rho_{\rm{c}}$~ and shallower density gradient in the
subsolar models relative to the solar models.

The dominant isotopes at C-depletion are $^{16}$O and $^{20}$Ne. These
two isotopes follows nearly Gaussian profiles with 95\% CI spreads
of $\simeq \pm 40\%$ and $\simeq \pm 70\%$ for $X_{\rm{c}}(^{16}$O)
and  $X_{\rm{c}}(^{20}$Ne), respectively. Despite this spread,
the zero variation values of $^{16}$O and $^{20}$Ne for
the solar and subsolar models are within $\simeq$~1\%.

\subsubsection{Spearman Correlation Coefficients}
\label{s.sroc_c}

Figure~\ref{fig:all_rhos_c_dep} shows the absolute SROCs
for the 665 sampled reaction rates for the eight quantities considered in
Figure~\ref{fig:all_pdfs_c_dep} for the solar and subsolar grid of models.

Competition between the $^{12}$C\,+$^{12}$C and $^{12}$C\,+$^{16}$O
reaction rates largely determines the mass of the ONe core at C-depletion.
The $^{12}$C($^{12}$C,$p$)$^{23}$Na rate have significant positive SROC
values of $r_{\rm{s}}=(+0.58,+0.56)$ for the solar and subsolar models,
respectively. Protons produced by $^{12}$C($^{12}$C,$p$)$^{23}$Na are
usually captured by $^{23}$Na(p,$\alpha$)$^{20}$Ne, which increases $M_{\rm{ONe-Core}}$.
Uncertainties in the $^{12}$C($^{16}$O,$p$)$^{27}$Al and
$^{12}$C($^{16}$O,$\alpha$)$^{27}$Mg rates have significant negative SROC 
values, $r_{\rm{s}}$\,$\simeq$\,$-0.40$, because the main products from these
reactions ultimately produce $^{28}$Si, which decreases $M_{\rm{ONe-Core}}$
by effectively transferring $^{16}$O to $^{28}$Si
\citep{woosley_1971_aa,martinez-rodriguez_2017_aa,fang_2017_aa}.

The $^{12}$C($\alpha,\gamma$)$^{16}$O rate impacts the time between
He-depletion and C-depletion $\tau_{\rm{C-burn}}$ with SROC values of
$\simeq$\,+0.91 and $\simeq$\,+0.94 for the solar and subsolar models,
respectively. This occurs because this rate sets the mass of the
CO core, which has a relatively narrow 95\% CI range of $\simeq
\pm~2\%$ (see Figure \ref{fig:all_pdfs_he_dep}).  Smaller uncertainties
in the triple-$\alpha$ rate (negative correlation) and the
$^{14}$N($p,\gamma$)$^{15}$O rate (positive correlation) occur because
these two reactions play a diminished role in setting the mass of
the CO core.

The SROC analysis for $T_{\rm{c}}$ and $\rho_{\rm{c}}$ shows
dependencies on the $^{12}$C($\alpha,\gamma$)$^{16}$O,
$^{12}$C\,+$^{12}$C, and $^{12}$C\,+$^{16}$O rates for the solar and
subsolar models.  All these rates have negative SROCs of $r_{\rm{s}}
\simeq$\,$-0.4$.  These magnitude and sign are partially due to
thermal neutrino losses playing a key role in the evolution, and
partially due to the thermostat mechanism, namely larger energy
producing reaction rates yield cooler and less dense cores.

The quantities $Y_{\rm{e,c}}$ and $\xi_{2.5}$ inherit a dependence on the
$^{12}$C($\alpha,\gamma$)$^{16}$O rate from He-burning, with SROCs
of $r_{\rm{s}}$\,$\simeq$\,(+0.7, +0.5), respectively.  Uncertainties in
the $^{12}$C($^{12}$C,$p$)$^{23}$Na and $^{12}$C($^{16}$O,$p$)$^{27}$Al
rates also contribute with negative coefficients.

Likewise, $X_{\rm{c}}(^{16}$O) and $X_{\rm{c}}(^{20}$Ne) also inherit
a strong dependence on the $^{12}$C($\alpha,\gamma$)$^{16}$O rate from
He-burning, with SROCs of $r_{\rm{s}}$~$\simeq$~(+0.9, $-$0.8),
respectively.  The $X_{\rm{c}}(^{16}$O) has a positive correlation
coefficient because during He-burning a larger
$^{12}$C($\alpha,\gamma$)$^{16}$O rate produces more $^{16}$O. The
$X_{\rm{c}}(^{20}$Ne) has a negative SROC
because a larger $^{12}$C($\alpha,\gamma$)$^{16}$O rate produces less
$^{12}$C, the principal fuel of C-burning, which produces less $^{20}$Ne.
Both isotopes also share a smaller dependency on the triple-$\alpha$ rate 
uncertainty, inherited from He-burning, and a small dependence on C-burning 
rates. These smaller dependencies are also anti-correlated $-$ increases in 
rates that increase $X_{\rm{c}}(^{16}$O) also decrease $X_{\rm{c}}(^{20}$Ne), 
and vice versa.

\subsubsection{Impact of the Measurement Point}
\label{s.meas_c}

The 95\% CI width of the $M_{\rm{ONe-Core}}$ PDF in Figure
\ref{fig:all_pdfs_c_dep} is partly due to the measurement point. The
$M_{\rm{ONe-Core}}$ is still growing in mass due to the off-center
convective C-burning episodes (See Figure~\ref{fig:baseline_kipps}).
This contrasts with H and He where convective core burning accounted 
for complete mixing of the ash of the nuclear burning.

In more detail, carbon ignites centrally and convectively in these 15\,\msun models.
The extent of the convective core burning reaches
$\simeq$\,0.6\,\msun.
Convection retreats as carbon is
depleted, and by $X_{\rm{c}}(^{12}\textup{C})$~$\simeq$~10$^{-2}$ the entire core
is radiative.  Subsequently, the first off-center convective C-burning
episode occurs when $X_{\rm{c}}(^{12}\textup{C})$~$\simeq$10$^{-4}$ and extends
from $\simeq$\,0.6\,\msun to 1.2$-$2.0\,\msun depending on the amount of
C fuel available from core He-burning.  It is the variability of the
location and extent of the off-center convective C-burning
episodes, which occurs {\it before} the 
measurement point of $X_{\rm{c}}(^{12}\textup{C}) \lesssim 1\times10^{-6}$,
that drives the 95\% CI spread in the $M_{\rm{ONe-Core}}$ PDF.

\begin{figure}[!htb]
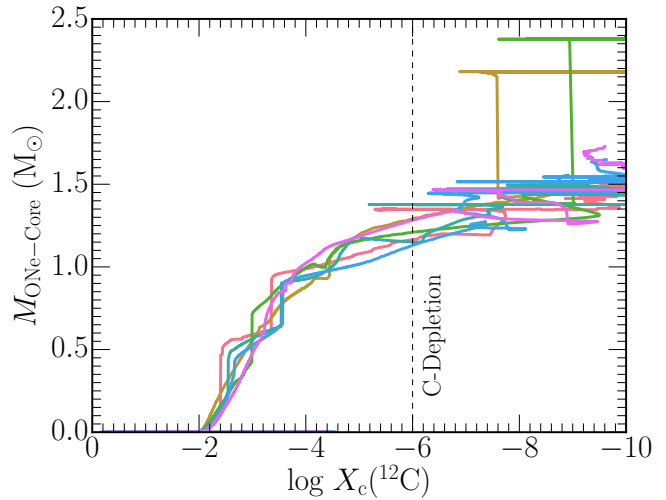

\centering{\includegraphics[width=\columnwidth]{{{ONe_core_mass_v_center_c12_solar-cropped}}}}
\caption{
Mass of the ONe core as a function of the central $^{12}$C mass fraction
for six solar grid models. Our adopted measurement point for C-depletion,
$X_{\rm{c}}(^{12}\textup{C}) \lesssim 1\times10^{-6}$, is shown by the dashed
vertical line. Variation in the mass of the ONe core is driven by the size and extent of
off-center convective C-burning episodes.
}
\label{fig:one_core}
\end{figure}

Figure~\ref{fig:one_core} shows the impact of the measurement point on
$M_{\rm{ONe-Core}}$ as a function of $X_{\rm{c}}(^{12}\textup{C})$ for
six solar grid models.  The dashed vertical line shows our
measurement point for C-depletion, $X_{\rm{c}}(^{12}\textup{C})
\lesssim 1\times10^{-6}$.  Given different compositions and
thermodynamic trajectories inherited from core He-burning, some models
are further along in transforming the CO core to a ONe core.  Despite
the 95\% CI range in the $M_{\rm{ONe-Core}}$ PDF, our SROC analysis yields
qualitatively similar results. Moreover, two models - the green and gold lines, 
grow larger ONe cores due to the extent of convective zone of the 
final off-center C burning episode mixing the fuel and ash of C-burning outward
to a larger mass coordinate than the remaining three models.

\subsection{Neon Depletion}

\begin{figure*}[!htb]
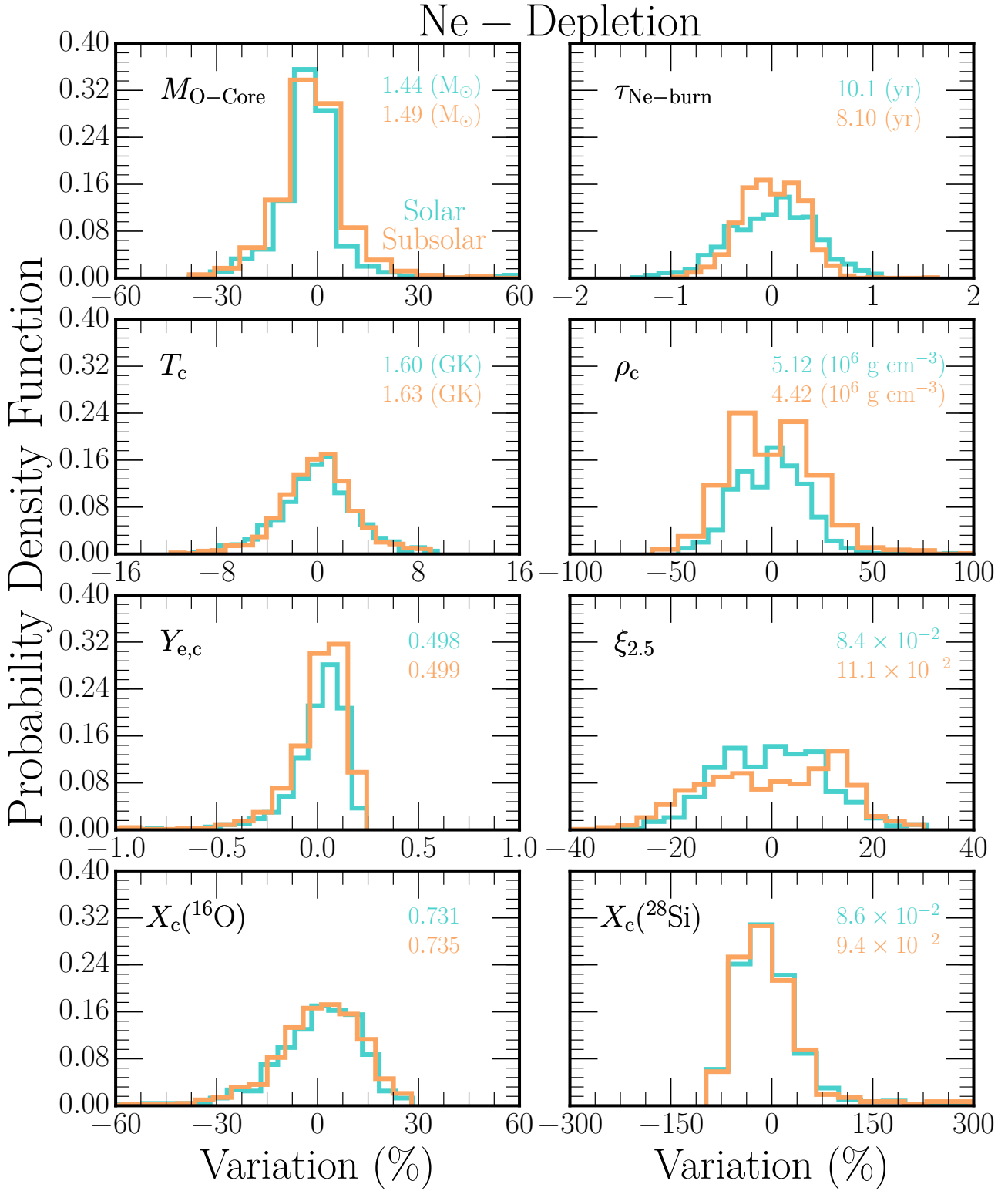

\centering{\includegraphics[width=2.0\columnwidth]{{{all_pdfs_ne_dep_v4-cropped}}}}
\caption{
Same as in Figure~\ref{fig:all_pdfs_c_dep} except we consider
$M_{\rm{O-Core}}$ - the mass of the O core,
$\tau_{\rm{Ne-burn}}$ - the elapsed time between C-depletion and Ne-depletion,
$T_{\rm{c}}$ - the central temperature,
$\rho_{\rm{c}}$ - the central density,
$Y_{\rm{e,c}}$ - the central electron fraction,
$\xi_{2.5}$ - the compactness parameter,
$X_{\rm{c}}(^{16}$O) - the central oxygen-16 mass fraction,
and $X_{\rm{c}}(^{28}$Si) - the central silicon-28 mass fraction.
All quantities are measured at Ne-depletion.
}
\label{fig:all_pdfs_ne_dep}
\end{figure*}

\begin{figure*}[!htb]
\centering{\includegraphics[width=2.0\columnwidth]{{{all_rhos_ne_dep_v5-cropped}}}}
\caption{
Same as in Figure~\ref{fig:all_rhos_c_dep}.
The quantities considered are
$M_{\rm{O-Core}}$ - the mass of the O core,
$\tau_{\rm{Ne-burn}}$ - the elapsed time between C-depletion and Ne-depletion,
$T_{\rm{c}}$ - the central temperature,
$\rho_{\rm{c}}$ - the central density,
$Y_{\rm{e,c}}$ - the central electron fraction,
$\xi_{2.5}$ - the compactness parameter,
$X_{\rm{c}}(^{16}$O) - the central oxygen-16 mass fraction,
and $X_{\rm{c}}(^{28}$Si) - the central silicon-28 mass fraction.
All quantities are measured at Ne-depletion.
}
\label{fig:all_rhos_ne_dep}
\end{figure*}

Core Ne-depletion is the next evolutionary stage considered.
We measure the integrated impact of the rate uncertainties
at the point when the central neon mass fraction
$X_{\rm{c}}(^{20}\textup{Ne}) \lesssim 1\times10^{-3}$.
This is a larger mass fraction than the $1\times10^{-6}$
used for H, He and C-depletion. We use a larger depletion value 
because a growing convective core feeds unburned neon
into the core. Ne does not deplete to $1\times10^{-6}$ until
well into core O-burning.

\subsubsection{Probability Distribution Functions}
\label{s.pdf_ne}

Figure~\ref{fig:all_pdfs_ne_dep}, shows the PDFs
of eight properties of the stellar models at Ne-depletion.
We consider
the mass of the O core $M_{\rm{O-Core}}$,
elapsed time between C-depletion and Ne-depletion $\tau_{\rm{Ne-burn}}$,
central temperature $T_{\rm{c}}$,
central density $\rho_{\rm{c}}$,
central electron fraction $Y_{\rm{e,c}}$,
compactness parameter $\xi_{2.5}$,
central oxygen-16 mass fraction $X_{\rm{c}}(^{16}$O), and
central silicon-28 mass fraction and $X_{\rm{c}}(^{28}$Si).

The $M_{\rm{O-Core}}$ PDF shows a strong peak for the
solar and subsolar models with zero variation values of
1.44\,\Msun and 1.49\,\Msun, respectively. The 95\% CI spread
is $\simeq \pm 30 \%$ for both sets of models. The peaks are offset from
zero due to the long tail of positive variations.
The $\tau_{\rm{Ne-burn}}$ PDFs show 95\% CIs of $\simeq \pm 1\%$,
commensurate with the $\tau_{\rm{C-burn}}$ in Figure \ref{fig:all_pdfs_c_dep}.
The 95\% CI spread of the solar grid is slightly larger than the spread
for the subsolar grid. Both PDFs are symmetric about zero variations
of 10.1 yr and 8.10 yr, respectively.

The $T_{\rm{c}}$ distribution has zero variation values of 1.60\,GK and
1.63\,GK for the solar and subsolar grids, respectively. Both PDFs
are symmetric about their zero variation values, and have
95\% CI widths of $\simeq \pm$~6\%.  The $\rho_{\rm{c}}$ PDF has zero variation
values of 5.12 $\times10^6$ g cm$^{-3}$ and 4.42 $\times10^6$ g cm$^{-3}$ for the 
solar and subsolar gridss, respectively. Both PDFs have 95\% CI widths
of $\simeq\,\pm$~50\%. The subsolar model PDF has a slight bimodality
with equal peaks of $\simeq$\,18\%.  The
$T_{\rm{c}}$ and $\rho_{\rm{c}}$ PDFs have 95\% spreads that are smaller than
the corresponding 95\% CI widths for C-depletion.

The $Y_{\rm{e,c}}$ PDFs for both metallicity grids strongly
peak about their means, 0.498 and 0.499 respectively, with a 95\% spread of
$\lesssim$~0.25\%. This is about the same 95\% CI spread as at
C-depletion, reflecting that significant neutronization does not occur
during Ne-burning. The $\xi_{2.5}$ PDF shows a 95\% CI spread 
of $\simeq$\,$\pm$\,20\% without a strong central peak for both metallicity grids.

$X_{\rm{c}}(^{16}$O) follows a broad distribution about the mean
with variations of $\simeq$\,(+20\%,-30\%). In contrast, $X_{\rm{c}}(^{28}$Si), 
the other dominant isotope at Ne-depletion, follows a more centrally 
peaked distribution but with a larger width of $\simeq -120\%$ and a 
slight, long tail showing variations out to $\simeq +200\%$ of the mean.

\subsubsection{Spearman Correlation Coefficients}
\label{s.sroc_ne}

Figure~\ref{fig:all_rhos_ne_dep} shows the SROC correlations 
for the eight quantities considered in Figure~\ref{fig:all_pdfs_ne_dep}.
Markers and colors are the same as in Figure~\ref{fig:all_rhos_c_dep}.

Ne-depletion inherits most of the reaction rate dependencies
from He-depletion and C-depletion. This is consistent with Ne-burning being
a photodisintegration rearrangement, whose net reaction
is 2($^{20}$Ne) $\rightarrow$ $^{16}$O  + $^{24}$Mg  + 4.6 Mev. The nucleosynthesis 
products also resemble those at C-depletion but lack $^{23}$Na and has more of the 
heavier nuclei $^{26,27}$Al, $^{29,30}$Si, and $^{31}$P.

The 95\% CI spread of $M_{\rm{O-Core}}$ is mainly driven by rate uncertainties
in $^{12}$C($^{12}$C,$p$)$^{23}$Na, with $r_{\rm{s}}\simeq+0.8$ for both metallicity grids. The $^{12}$C($\alpha,\gamma$)$^{16}$O
rate also affects the O core mass but to a lesser extent, with $r_{\rm{s}}\simeq+0.4$.
The 95\% CI variation of $\tau_{\rm{Ne-burn}}$ follows that of
the spread of $\tau_{\rm{C-burn}}$. It is affected primarily by
uncertainties in the $^{12}$C($\alpha,\gamma$)$^{16}$O rate with
smaller dependencies on rate uncertainties in
$^{14}$N($p,\gamma$)$^{15}$O (positive SROC) and 
triple-$\alpha$ (negative SROC).
In general, the SROC values are larger for the solar grid.

\begin{figure}[!htb]
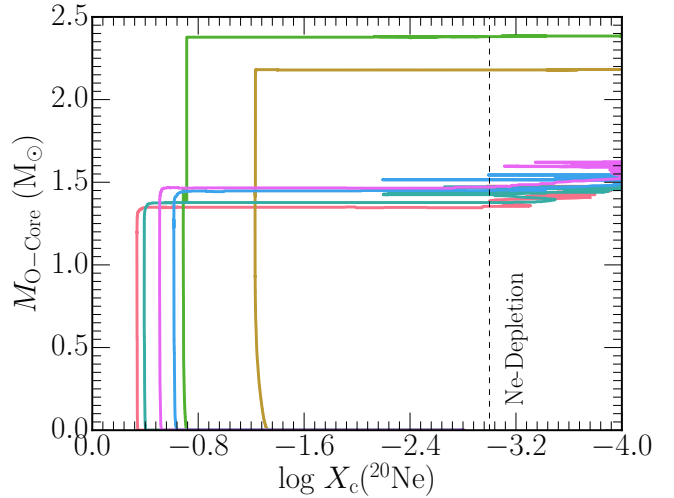

\centering{\includegraphics[width=\columnwidth]{{{O_core_mass_v_center_ne20_solar-cropped}}}}
\caption{
Mass of the O core as a function of the central $^{20}$Ne mass fraction
for the same six solar grid models from Figure~\ref{fig:one_core}.
Our adopted measurement point for Ne-depletion,
$X_{\rm{c}}(^{20}\textup{Ne}) \lesssim 1\times10^{-3}$, is shown by the dashed
vertical line. Variation in the mass of the O core is driven by the extent of the
final off-center C flash prior to Ne-depletion.
}
\label{fig:ne_dep_o_core_mass_v_ne}
\end{figure}

The $T_{\rm{c}}$ PDF depends mostly on the uncertainties in the 
$^{12}$C($\alpha,\gamma$)$^{16}$O rate
for both solar and subsolar grids. The positive SROC implies that a
larger $^{12}$C($\alpha,\gamma$)$^{16}$O rate yields a hotter stellar core.
This is the first occurrence of an inversion of the thermostat mechanism.
A small dependency is also found for triple-$\alpha$ and the 
$^{12}$C($^{12}$C,$p$)$^{23}$Na rates. 

The central density has an SROC value of $r_{\rm{s}}$\,$\simeq$\,+0.5 for the
$^{12}$C($^{12}$C,$p$)$^{23}$Na rate. However, $\rho_{\rm{c}}$ is also
affected by uncertainties in the $^{16}$O($^{16}$O,$p$)$^{31}$P rate with
$r_{\rm{s}}$\,$\simeq$\,$-0.3$. This indicates O-burning is beginning to take
place at Ne-depletion. There is also a weaker dependence on the
$^{16}$O($^{16}$O,$^4$He)$^{28}$Si rate with a negative SROC.

Uncertainties in the $^{12}$C($\alpha,\gamma$)$^{16}$O rate drive the variations in
$Y_{\rm{e,c}}$ and $\xi_{2.5}$ with SROC values of $r_{\rm{s}}$\,$\simeq$\,+0.6.
Smaller SROCs are also found for $Y_{\rm{e,c}}$ and the 95\% CI
$^{27}$Al($\alpha,p$)$^{30}$Si,
triple-$\alpha$,
$^{16}$O($^{16}$O,$p$)$^{31}$P,
$^{16}$O($^{16}$O,$n$)$^{31}$S, and
$^{12}$C($^{12}$C,$p$)$^{23}$Na rates.
Similar to $\rho_{\rm{c}}$,  the compactness of the stellar core is
weakly affected by the inherited uncertainties from the $^{12}$C($^{12}$C,$p$)$^{23}$Na rate.

$X_{\rm{c}}(^{16}$O) inherits a dependence on the
$^{12}$C($\alpha,\gamma$)$^{16}$O rate with $r_{\rm{s}}$\,$\simeq$\,+0.7
for both metallicity grids. The $^{12}$C\,+$^{12}$C and $^{16}$O\,+$^{16}$O rates
have smaller, competing affects on $X_{\rm{c}}(^{16}$O)
with $\left | r_{\rm{s}} \right | \simeq0.25$.
$X_{\rm{c}}(^{28}$Si) is slightly anti-correlated with $X_{\rm{c}}(^{16}$O),
with the $^{12}$C($\alpha,\gamma$)$^{16}$O rate having the largest (negative) SROC.
Smaller effects from the uncertainties in the heavy ion, carbon and
oxygen channels also play a role in its variation.

\newpage

\subsubsection{Impact of the Measurement Point}
\label{s.meas_ne}

Some of the quantities measured at Ne-depletion partly
inherit their 95\% CI spread from the spread at C-depletion.
However, the spread of most quantities at  Ne-depletion is larger
than the 95\% CI spreads at C-depletion because of the
thermodynamic conditions imposed by the depletion of carbon.

Figure~\ref{fig:one_core} shows the extent of the ONe core, measured
at C-depletion is sensitive to the extent of the final off-center
convective carbon episode.
Figure~\ref{fig:ne_dep_o_core_mass_v_ne} shows $M_{\rm{O-Core}}$
as a function of the central $^{20}$Ne mass fraction
for the same six solar grid models as in Figure~\ref{fig:one_core}.
The same two stellar models which yield larger ONe core masses
in Figure~\ref{fig:one_core}, introduce larger 95\% CI variations
in the O core mass measured at Ne-depletion.
The variation in the ONe core mass inherited from
C-depletion can cause variations in the other measured quantities.
We stress that our analysis measures the integrated impact of the reaction rate
uncertainties on the evolution of the stellar model up to the measurement point.

\subsection{Oxygen Depletion}

\begin{figure*}[!htb]
\centering{\includegraphics[width=2.0\columnwidth]{{{all_pdfs_o_dep_v5-cropped}}}}
\caption{
Same as in Figure~\ref{fig:all_pdfs_ne_dep} except we consider
$M_{\rm{Si-Core}}$ - the mass of the Si core,
$\tau_{\rm{O-burn}}$ - the elapsed time between Ne-depletion and O-depletion,
$T_{\rm{c}}$ - the central temperature,
$\rho_{\rm{c}}$ - the central density,
$Y_{\rm{e,c}}$ - the central electron fraction,
$\xi_{2.5}$ - the compactness parameter,
$X_{\rm{c}}(^{28}$Si) - the central silicon-28 mass fraction, and
$X_{\rm{c}}(^{32}$S) - the central sulfur-32 mass fraction all measured at O-depletion.}
\label{fig:all_pdfs_o_dep}
\end{figure*}

The last evolutionary point we consider is core O-depletion, defined when
$X_{\rm{c}}(^{16}\textup{O}) \lesssim 1\times10^{-3}$.
We consider eight properties of the stellar model at this epoch:
mass of the Si core $M_{\rm{Si-Core}}$,
time between Ne-depletion and O-depletion $\tau_{\rm{O-burn}}$,
central temperature $T_{\rm{c}}$,
central density $\rho_{\rm{c}}$,
central electron fraction $Y_{\rm{e,c}}$,
compactness parameter $\xi_{2.5}$,
central silicon-28 mass fraction $X_{\rm{c}}(^{28}$Si) , and
central sulfur-32 mass fraction $X_{\rm{c}}(^{32}$S).

\subsubsection{Probability Distribution Functions}
\label{s.pdf_o}

Figure~\ref{fig:all_pdfs_o_dep} shows the variation of these
quantities in the same format as for previous depletion epochs.
The $M_{\rm{Si-Core}}$ PDF for the solar models span $\simeq$\,$-$120
to $+$400.  Only the range $\pm$\,120 is shown in
Figure~\ref{fig:all_pdfs_o_dep}.  The full range, which is taken into
account in the analysis, causes the peak to center at
$\simeq$\,$-$30\%. Despite the wide range, the zero variation values
of 0.27\,\Msun for the solar grid and 0.22\,\Msun for the
subsolar grid are similar.  The 95\% CI spreads are $\approx$\,4 times
larger for O-depletion than for Ne-depletion for both the solar and
subsolar grids.

The solar and subsolar $\tau_{\rm{O-burn}}$ PDFs have zero variation
values of 3.79 yr and 2.35 yr, respectively. The 95\% CI spreads of $\simeq$\,$\pm$~1\%
are consistent with the  95\% CI  lifetimes of previous epochs.
The subsolar model PDF has a slightly larger peak amplitude and smaller range.

The solar and subsolar $T_{\rm{c}}$ PDFs have a 95\% CI width of
$\simeq$\,$\pm$\,10\%.  The negative variation tail causes a $\simeq$\,$-$20\%
shift away from the zero variation values of 2.07~GK
for the solar models and 2.14~GK for the subsolar models.
The $\rho_{\rm{c}}$ PDFs have 95\% CI spreads of $\simeq$\,$\pm$\,60\% with
tails out to $\simeq$\,+160\% for both metallicities.  These tails
cause the peak in the PDF to shift away from the arithmetic means of 
23.3$\times$10$^6$~g~cm$^{-3}$
for the solar models and 15.1$\times$10$^6$~g~cm$^{-3}$
for the subsolar models. Commensurate with Figure \ref{fig:baseline_TRho},
the solar models remain cooler and denser than the subsolar
models at O-depletion.

At the elevated $T_{\rm{c}}$ and $\rho_{\rm{c}}$ that occurs during O-depletion, the
reactions $^{16}$O($^{16}$O,$n$)$^{31}$S, $^{33}$S(e$^-$,$\nu$)$^{33}$P,
$^{37}$Ar(e$^-$,$\nu$)$^{33}$Cl, and $^{35}$Cl(e$^-$,$\nu$)$^{35}$S
decrease $Y_{\rm{e,c}}$. This is reflected in the $Y_{\rm{e,c}}$ PDF
having zero variation values of 0.492 and 0.493 for the solar and subsolar 
grids, respectively. Peaks in the $Y_{\rm{e,c}}$ PDF are shifted from these
zero variation because both the solar and subsolar grids have
tails of negative variations extending to $\simeq$\,$-$2\%.

The $\xi_{2.5}$ PDFs show 95\% CI spreads of $\simeq$\,$\pm$\,20\% for
the solar and subsolar grids. The arithmetic means of
$\xi_{2.5}$\,$\simeq$\,0.102 for the solar grid and
$\xi_{2.5}$\,$\simeq$\,0.139 for the subsolar grid are larger than
the arithmetic means at Ne-depletion, but the difference in
$\xi_{2.5}$ between the two metallicities are similar.
The $X_{\rm{c}}(^{28}$Si) PDF is log-normal
with a peak at $\simeq$\,-45\% with extrema extending to variations of
$\simeq$\,$-120$\% and $\simeq$\,+180\%. The $X_{\rm{c}}(^{32}$S) PDF
is broad with tails extending to $\simeq$\,$\pm$\,80\%.

The 95\% CI widths of the PDFs for $M_{\rm{Si-Core}}$ is driven by 
the fact that the Si-core is still forming at the measurement point of
$X_{\rm{c}}(^{16}\textup{O}) \lesssim 1\times10^{-3}$.
Additional dynamic range is introduced by
some models forming heavier isotopes of Si and S, and \MESA only
considering $^{28}$Si in the definition of the Si-core mass boundary.
For example, the central composition at O-depletion
for one of the models in Figure~\ref{fig:one_core}
and \ref{fig:ne_dep_o_core_mass_v_ne} is
$X_{\rm{c}}(^{28}\rm{Si})$\,$\simeq$\,$4.6\times10^{-2}$,
$X_{\rm{c}}(^{30}\rm{Si})$\,$\simeq$\,$3.5\times10^{-1}$,
$X_{\rm{c}}(^{32}\rm{S})$\,$\simeq$\,$4.6\times10^{-2}$,
and
$X_{\rm{c}}(^{34}\rm{S})$\,$\simeq$\,$4.4\times10^{-1}$.
This model reports a very small  $M_{\rm{Si-Core}}$
because Si is primarily in the neutron rich $^{30}\rm{Si}$.
This also accounts for the negative tail in the $Y_{\rm{e,c}}$ PDF and
the dynamic range in $\rho_{\rm{c}}$.

\begin{figure*}[!htb]
\centering{\includegraphics[width=2.0\columnwidth]{{{all_rhos_o_dep_v5-cropped}}}}
\caption{
Same as in Figure~\ref{fig:all_rhos_ne_dep}.
The quantities considered are
$M_{\rm{Si-Core}}$ - the mass of the Si core,
$\tau_{\rm{O-burn}}$ - the elapsed time between Ne-depletion and O-depletion,
$T_{\rm{c}}$ - the central temperature,
$\rho_{\rm{c}}$ - the central density,
$Y_{\rm{e,c}}$ - the central electron fraction,
$\xi_{2.5}$ - the compactness parameter,
$X_{\rm{c}}(^{28}$Si) - the central silicon-28 mass fraction, and
$X_{\rm{c}}(^{32}$S) - the central sulfur-32 mass fraction.
All quantities are measured at O-depletion.}
\label{fig:all_rhos_o_dep}
\end{figure*}

\subsubsection{Spearman Correlation Coefficients}
\label{s.sroc_o}

Figure~\ref{fig:all_rhos_o_dep} shows the SROC
coefficients for the solar and subsolar grid against the eight 
quantities in Figure~\ref{fig:all_pdfs_o_dep}.
The format is the same as in previous figures.

The $M_{\rm{Si-Core}}$ has a negative correlation of $r_{\rm{s}}$\,$\simeq$\,$-0.25$
with the $^{16}$O($^{16}$O,$\alpha$)$^{28}$Si rate and a smaller dependence
on the $^{12}$C($\alpha,\gamma$)$^{16}$O rate.  Reaction rates whose
uncertainty most impacts $\tau_{\rm{O-burn}}$ are inherited from
previous stages, namely $^{12}$C($\alpha,\gamma$)$^{16}$O, 
$^{14}$N($p,\gamma$)$^{15}$O, and triple-$\alpha$.

$T_{\rm{c}}$ has a negative correlation with the
$^{16}$O($^{16}$O,$n$)$^{31}$S rate for the solar and subsolar grid.
$\rho_{\rm{c}}$ inherits its dependence on the $^{12}$C($\alpha,\gamma$)$^{16}$O rate with
$r_{\rm{s}}\simeq-0.45$.  The other $^{16}$O\,+\,$^{16}$O exit channels have smaller effects
on $T_{\rm{c}}$ and $\rho_{\rm{c}}$.
The $^{16}$O($^{16}$O,$n$)$^{31}$S rate dominate the
SROCs for $Y_{\rm{e,c}}$ with $r_{\rm{s}}$\,$\simeq$\,$-0.6$ for both grids.
$\xi_{2.5}$ inherits dependencies on
the $^{12}$C($\alpha,\gamma$)$^{16}$O and $^{12}$C($^{12}$C,$p$)$^{23}$Na rates.
The mass fractions $X_{\rm{c}}(^{28}$Si) and $X_{\rm{c}}(^{32}$S)
are chiefly the result of the competition between
the $^{16}$O($^{16}$O,$\alpha$)$^{28}$Si and
$^{28}$Si($\alpha,\gamma$)$^{32}$S rates.

Table \ref{tbl:all_properties} summarizes the properties of the PDFs and the SROC analysis
at O-depletion, along with the results for previous depletion points of the major fuels. 

\begin{deluxetable*}{cccccccc}
\renewcommand{\arraystretch}{1.05}
\tablecolumns{8}
\tablewidth{2.0\linewidth}
\tablecaption{Properties of 15\,M$_{\odot}$ Solar and Subsolar Monte Carlo Stellar Models}
\tablehead{
\multicolumn{1}{c}{} & \multicolumn{3}{c}{Solar} & \multicolumn{3}{c}{Subsolar} & \multicolumn{1}{c}{} \\
\hline
\hline
\colhead{Property} &
\colhead{Values} &  \colhead{Key Reaction} & \colhead{$r_{\rm{s}}$} &
\colhead{Values} &  \colhead{Key Reaction} & \colhead{$r_{\rm{s}}$} &
\colhead{95\% CI Limits of Variation~(\%)} \\
\hline
\multicolumn{8}{c}{H-Depletion}
}
\startdata
$M_{\rm{He-Core}}~(\rm{M}_{\odot})$ &
2.802$^{2.806}_{2.799}$ & $^{14}$N($p,\gamma$)$^{15}$O & +0.35 &
2.86$^{2.863}_{2.857}$ & $\ldots$ & +0.33 &
(+0.13, -0.12)~(+0.12,-0.12)  \\
$\tau_{\rm{TAMS}}~(\rm{Myr})$ &
11.27$^{11.29}_{11.249}$ & $^{14}$N($p,\gamma$)$^{15}$O & +1.0 &
11.769$^{11.78}_{11.759}$ & $\ldots$ & +1.0 &
(+0.18,-0.19)~(+0.09,-0.09)  \\
$T_{\rm{c}}~(\rm{10^{8}}~K)$ &
0.628$^{0.635}_{0.621}$ & $^{14}$N($p,\gamma$)$^{15}$O & -0.99 &
0.807$^{0.816}_{0.799}$ & $\ldots$ & -0.99 &
(+1.19,-1.12)~(+1.07,-1.11)  \\
$\rho_{\rm{c}}~(\rm{g~cm}^{-3})$ &
42.402$^{43.917}_{40.978}$ & $^{14}$N($p,\gamma$)$^{15}$O & -0.97 &
88.198$^{91.241}_{85.122}$ & $\ldots$ & -0.98 &
(+3.57,-3.36)~(+3.45,-3.49)  \\
$\xi_{\rm{2.5}}$ &
0.007$^{0.007}_{0.007}$ & $^{14}$N($p,\gamma$)$^{15}$O & -0.99 &
0.008$^{0.009}_{0.008}$ & $\ldots$ & -0.99  &
(+1.19,-1.14)~(+1.09,-1.13)  \\
$X_{\rm{c}}(^{14}\rm{N})\times10^{3}$ &
9.234$^{9.317}_{9.128}$ &  $^{15}$N($p,\gamma$)$^{16}$O & -0.56 &
0.194$^{0.195}_{0.192}$ &  $^{14}$N($p,\gamma$)$^{15}$O & -0.66 &
(+0.89,-1.15)~(+0.83,-1.08)  \\
\hline
\multicolumn{8}{c}{He-Depletion} \\
\hline
$M_{\rm{CO-Core}}~(\rm{M}_{\odot})$ &
2.414$^{2.522}_{2.347}$ & $^{12}$C($\alpha,\gamma$)$^{16}$O & +0.79 &
2.952$^{3.062}_{2.909}$ & $\ldots$ & +0.84 &
(+1.94,-3.10)~(+1.53,-2.44)  \\
$\tau_{\rm{He-burn}}~(\rm{Myr})$ &
1.594$^{1.696}_{1.479}$ & $^{12}$C($\alpha,\gamma$)$^{16}$O & +0.92 &
1.315$^{1.393}_{1.234}$ & $\ldots$ & +0.94 &
(+0.81,-0.90)~(+0.58,-0.63)  \\
$T_{\rm{c}}~(\rm{10^{8}}~K)$ &
3.126$^{3.185}_{3.092}$ & Triple-$\alpha$ & -0.80 &
3.207$^{3.26}_{3.171}$ & $\ldots$ & -0.77 &
(+1.89,-1.07)~(+1.67,-1.11)  \\
$\rho_{\rm{c}}~(10^3~\rm{g~cm}^{-3})$ &
5.535$^{5.756}_{5.364}$ & Triple-$\alpha$ & -0.69 &
5.053$^{5.265}_{4.892}$ & $\ldots$ & -0.67 &
(+3.99,-3.08)~(+4.19,-3.20)  \\
$X_{\rm{c}}(^{22}\rm{Ne})\times10^{2}$ &
1.081$^{1.217}_{0.887}$ & $^{22}$Ne($\alpha,\gamma$)$^{26}$Mg & -0.70 &
0.021$^{0.024}_{0.017}$ & $\ldots$ & -0.64 &
(+12.6,-17.9)~(+15.5,-21.5)  \\
$\xi_{\rm{2.5}}$ &
0.031$^{0.031}_{0.03}$ & Triple-$\alpha$ & -0.83 &
0.031$^{0.032}_{0.031}$ & $\ldots$ & -0.74 &
(+1.56,-0.97)~(+1.41,-1.06)  \\
$X_{\rm{c}}(^{12}\rm{C})$ &
0.260$^{0.467}_{0.077}$ & $^{12}$C($\alpha,\gamma$)$^{16}$O & -0.95 &
0.265$^{0.459}_{0.082}$ & $\ldots$ & -0.95 &
(+79.9,-70.2)~(+73.1,-69.1)  \\
$X_{\rm{c}}(^{16}\rm{O})$ &
0.716$^{0.896}_{0.512}$ & $^{12}$C($\alpha,\gamma$)$^{16}$O & +0.95 &
0.731$^{0.910}_{0.538}$ & $\ldots$ & +0.95 &
(+25.1,-28.5)~(+24.5,-26.3)  \\
\hline
\multicolumn{8}{c}{C-Depletion} \\
\hline
$M_{\rm{ONe-Core}}~(\rm{M}_{\odot})$ &
1.110$^{1.365}_{0.550}$ & $^{12}$C($^{12}$C,$p$)$^{23}$Na & +0.58 &
1.175$^{1.444}_{0.575}$ & $\ldots$ & +0.53 &
(+22.9,-50.5)~(+22.9,-51.1)  \\
$\tau_{\rm{C-burn}}~(\rm{kyr})$ &
30.74$^{41.87}_{26.51}$ & $^{12}$C($\alpha,\gamma$)$^{16}$O & +0.91 &
23.75$^{32.95}_{21.00}$ & $\ldots$ & +0.94 &
(+0.78,-0.86)~(+0.56,-0.62)  \\
$T_{\rm{c}}~(\rm{GK})$ &
1.158$^{1.412}_{1.025}$ & $^{12}$C($^{16}$O,$p$)$^{27}$Al & -0.45 &
1.196$^{1.598}_{1.034}$ & $^{12}$C($^{12}$C,$p$)$^{23}$Na & -0.50 &
(+21.9,-11.5)~(+33.6,-13.6)  \\
$\rho_{\rm{c}}~(10^6~\rm{g~cm}^{-3})$ &
5.317$^{8.815}_{2.711}$ & $^{12}$C($^{16}$O,$p$)$^{27}$Al & -0.43 &
4.371$^{7.285}_{1.978}$ & $\ldots$ & -0.43 &
(+65.8,-49.0)~(+66.7,-54.8)  \\
$Y_{\rm{e,c}}$ &
0.498$^{0.499}_{0.497}$ & $^{12}$C($\alpha,\gamma$)$^{16}$O & +0.72 &
0.498$^{0.500}_{0.497}$ & $\ldots$ & +0.71 &
(+0.15,-0.35)~(+0.15,-0.37)  \\
$\xi_{\rm{2.5}}$ &
0.083$^{0.095}_{0.069}$ & $^{12}$C($\alpha,\gamma$)$^{16}$O & +0.49 &
0.109$^{0.129}_{0.088}$ & $^{12}$C($^{12}$C,$p$)$^{23}$Na & -0.54 &
(+14.28,-16.6)~(+17.9,-19.4)  \\
$X_{\rm{c}}(^{16}\rm{O})$ &
0.622$^{0.861}_{0.373}$ & $^{12}$C($\alpha,\gamma$)$^{16}$O & +0.91 &
0.625$^{0.869}_{0.369}$ & $\ldots$ & +0.88 &
(+38.4,-40.0)~(+38.9,-41.0)  \\
$X_{\rm{c}}(^{20}\rm{Ne})$ &
0.266$^{0.459}_{0.072}$ & $^{12}$C($\alpha,\gamma$)$^{16}$O & -0.79 &
0.280$^{0.474}_{0.043}$ & $\ldots$ & -0.75 &
(+72.4,-73.0)~(+69.1,-84.5)  \\
\hline
\multicolumn{8}{c}{Ne-Depletion} \\
\hline
$M_{\rm{O-Core}}~(\rm{M}_{\odot})$ &
1.439$^{2.368}_{1.113}$ & $^{12}$C($^{12}$C,$p$)$^{23}$Na & +0.82 &
1.493$^{1.965}_{1.157}$ & $\ldots$ & +0.79 &
(+64.5,-22.7)~(+31.6,-22.5)  \\
$\tau_{\rm{Ne-burn}}~(\rm{yr})$ &
10.114$^{44.452}_{0.493}$ & $^{12}$C($\alpha,\gamma$)$^{16}$O & +0.91 &
8.103$^{37.39}_{0.143}$ & $\ldots$ & +0.94 &
(+0.78,-0.86)~(+0.56,-0.62)  \\
$T_{\rm{c}}~(\rm{GK})$ &
1.603$^{1.702}_{1.501}$ & $^{12}$C($\alpha,\gamma$)$^{16}$O & +0.80 &
1.626$^{1.727}_{1.520}$ & $\ldots$ & +0.72 &
(+6.15,-6.40)~(+6.21,-6.52)  \\
$\rho_{\rm{c}}~(10^6~\rm{g~cm}^{-3})$ &
5.119$^{6.986}_{3.485}$ & $^{12}$C($^{12}$C,$p$)$^{23}$Na & +0.50 &
4.422$^{6.322}_{2.770}$ & $\ldots$ & +0.52 &
(+36.5,-31.9)~(+43.0,-37.4)  \\
$Y_{\rm{e,c}}$ &
0.498$^{0.499}_{0.496}$ & $^{12}$C($\alpha,\gamma$)$^{16}$O & +0.59 &
0.499$^{0.500}_{0.497}$ & $\ldots$ & +0.60 &
(+0.19,-0.41)~(+0.19,-0.41)  \\
$\xi_{\rm{2.5}}$ &
0.084$^{0.101}_{0.068}$ & $^{12}$C($\alpha,\gamma$)$^{16}$O & +0.71 &
0.111$^{0.136}_{0.084}$ & $^{12}$C($^{12}$C,$p$)$^{23}$Na & -0.57 &
(+20.2,-19.3)~(+22.5,-24.4)  \\
$X_{\rm{c}}(^{16}\rm{O})$ &
0.731$^{0.888}_{0.503}$ & $^{12}$C($\alpha,\gamma$)$^{16}$O & +0.74 &
0.735$^{0.893}_{0.501}$ & $\ldots$ & +0.70 &
(+21.4,-31.2)~(+21.4,-31.8)  \\
$X_{\rm{c}}(^{28}\rm{Si})$ &
0.086$^{0.240}_{0.021}$ & $^{12}$C($\alpha,\gamma$)$^{16}$O & -0.62 &
0.094$^{0.296}_{0.018}$ & $\ldots$ & -0.59 &
(+179,-75.2)~(+216,-81.1)  \\
\hline
\multicolumn{8}{c}{O-Depletion} \\
\hline
$M_{\rm{Si-Core}}~(\rm{M}_{\odot})$ &
0.270$^{0.979}_{0.012}$ & $^{16}$O($^{16}$O,$\alpha$)$^{28}$Si & +0.32 &
0.219$^{0.965}_{0.013}$ & $\ldots$ & +0.21 &
(+262,-95.7)~(+341,-94.2)  \\
$\tau_{\rm{O-burn}}~(\rm{yr})$ &
3.786$^{7.574}_{1.554}$ & $^{12}$C($\alpha,\gamma$)$^{16}$O & +0.91 &
2.348$^{5.546}_{0.857}$ & $\ldots$ & +0.93 &
(+0.76,-0.88)~(+0.52,-0.61)  \\
$T_{\rm{c}}~(\rm{GK})$ &
2.073$^{2.278}_{1.793}$ & $^{16}$O($^{16}$O,$p$)$^{31}$P & -0.56 &
2.141$^{2.350}_{1.860}$ & $\ldots$ & -0.47 &
(+9.92,-13.5)~(+9.77,-13.2)  \\
$\rho_{\rm{c}}~(10^6~\rm{g~cm}^{-3})$ &
23.34$^{54.06}_{10.32}$ & $^{12}$C($\alpha,\gamma$)$^{16}$O & -0.42 &
15.10$^{43.06}_{7.140}$ & $\ldots$ & -0.38 &
(+132,-55.8)~(+185,-52.7)  \\
$Y_{\rm{e,c}}$ &
0.492$^{0.498}_{0.479}$ & $^{16}$O($^{16}$O,$n$)$^{31}$S & -0.67 &
0.493$^{0.499}_{0.479}$ & $\ldots$ & -0.71 &
(+1.18,-2.74)~(+1.18,-2.88)  \\
$\xi_{\rm{2.5}}$ &
0.102$^{0.122}_{0.075}$ & $^{12}$C($\alpha,\gamma$)$^{16}$O & +0.69 &
0.139$^{0.174}_{0.106}$ & $\ldots$ & +0.51 &
(+19.9,-27.0)~(+24.98,-23.6)  \\
$X_{\rm{c}}(^{28}\rm{Si})$ &
0.263$^{0.526}_{0.106}$ & $^{16}$O($^{16}$O,$\alpha$)$^{28}$Si  & +0.39 &
0.268$^{0.544}_{0.111}$ & $^{16}$O($^{16}$O,$p$)$^{31}$P & -0.40 &
(+99.7,-59.8)~(+103,-58.6)  \\
$X_{\rm{c}}(^{32}\rm{S})$ &
0.433$^{0.736}_{0.058}$ & $^{16}$O($^{16}$O,$p$)$^{31}$P & +0.61 &
0.449$^{0.778}_{0.067}$ & $\ldots$ & +0.59 &
(+70.1,-86.7)~(+73.1,-85.1)  \\
\enddata
\tablecomments{
Properties of the 15 $M_{\odot}$ solar and subsolar
stellar models at five different epochs.
The values given are arithmetic means, with upper and
lower limits from the 95\% CI.
Also listed are the min. or max. SROC coefficient values and the 
corresponding key nuclear reaction. The last column are the limits 
of the 95\% CI for the variations for the solar (left) and subsolar (right)
stellar models. Ellipses indicate the variation of a given quantity
for the subsolar models is dominated by the same key reaction as the 
solar models.}
\label{tbl:all_properties}
\end{deluxetable*}


\begin{figure*}[!htb]
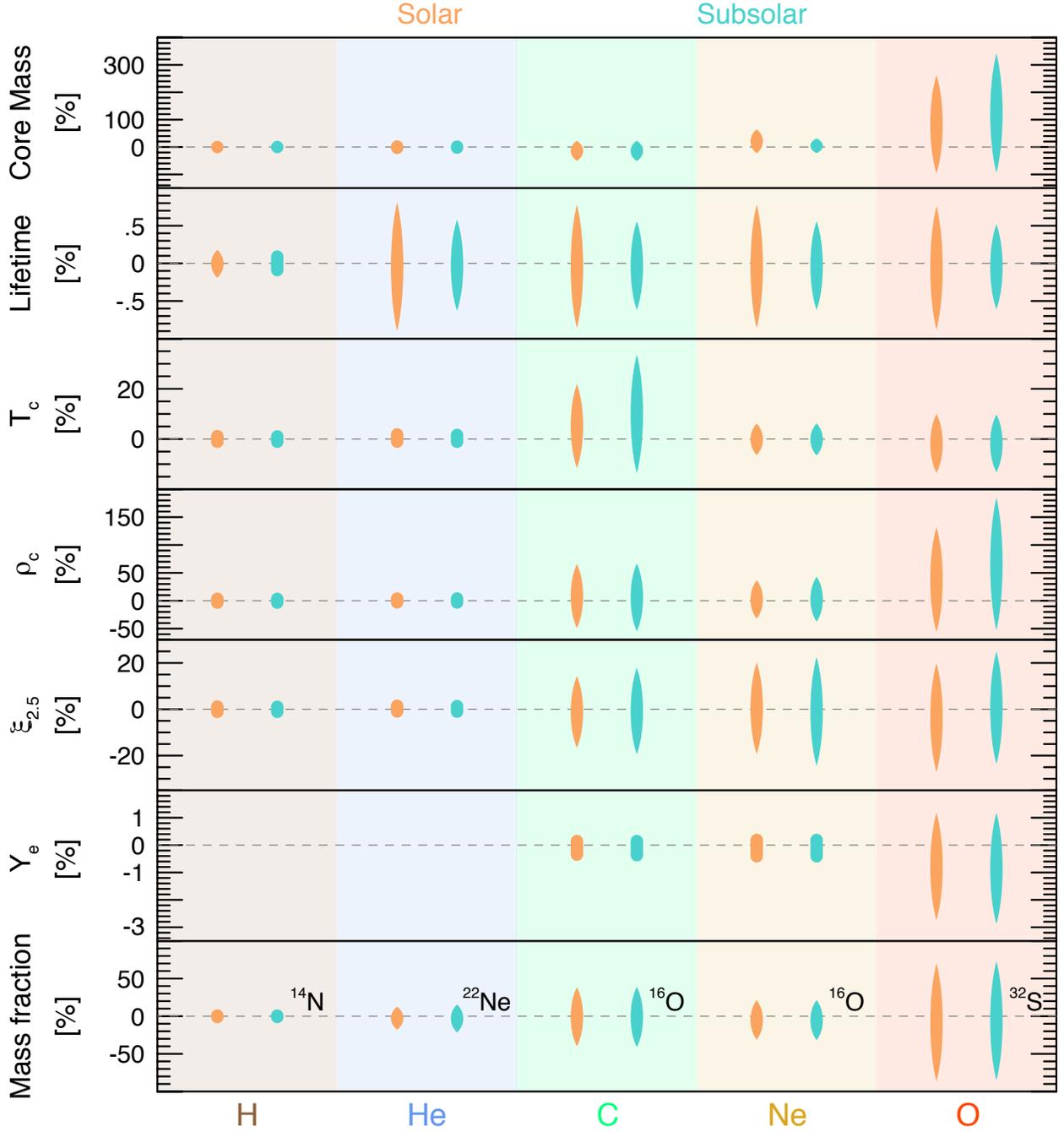

\centering{\includegraphics[width=2\columnwidth]{{{variations_msmc}}}}
\caption{
Percent variations of the core mass, lifetime, central temperature, 
central density, compactness parameter, electron fraction, and a 
chosen ``interesting'' mass fraction (top to bottom) at H-, He-, C-, Ne-, and O-depletion (left to right).
The vertical length of each tapered uncertainty band is the 95\% CI for variations 
about the mean arithmetic value, listed in the last column of 
Table \ref{tbl:all_properties}, and the horizontal width of each tapered uncertainty band 
schematically represents the underlying PDF. Solar metallicity models are shown by 
the orange bands and subsolar metallicity models by the green bands. 
The first occurrence of significant variation in
the compactness parameter $\xi_{2.5}$ occurs at C-depletion.
For the mass fractions,
we choose to show 
X$_{\rm c}$($^{14}$N) at H-depletion as it holds the star's initial CNO abundances,
X$_{\rm c}$($^{22}$Ne) at He-depletion as it holds the neutronization of the core,
X$_{\rm c}$($^{16}$O) at C- and Ne-depletion as it is dominant nucleosynthesis product of massive stars, and
X$_{\rm c}$($^{32}$S) at O-depletion as it is a key component of the ashes of O-burning.
}
\label{fig:variations_msmc}
\end{figure*}

\section{Discussion}
\label{sec:discussion}

Figure \ref{fig:variations_msmc} shows the 95\% CI variations listed
in Table~\ref{tbl:all_properties} for seven properties across five
evolutionary epochs for the solar and subsolar grids.  Across these
properties, the magnitude of the 95\% CI spreads generally grow with
each successive stage of evolution. The variations grow for two
reasons.  One, each evolutionary stage inherits variations from the
previous evolutionary stage because we measure the integrated impact
of the uncertainties in the reaction rates.  Two, each stage imprints
its own contributions to the variations due to the uncertainties in
the specific reaction rates that impact that stage. Finally, there is
a trend for the 95\% CI variations of the subsolar models to be
smaller than the variations of the solar models, particularly for
measurements at H-, He- and C-depletion.

We next discuss the reaction rates identified in Table~\ref{tbl:all_properties}
which have the largest impact on the variations of the core mass, burning lifetime, 
composition, and structural properties.

\subsection{Key Reaction Rates}

At H-depletion, Table~\ref{tbl:all_properties} shows the uncertainties from
the $^{14}$N($p,\gamma$)$^{15}$O reaction rate cause 95\% CI variations of
$\approx \pm$ 0.1\% in $M_{\rm{He-Core}}$,
$\approx \pm$ 0.2\% in $\tau_{\rm{TAMS}}$,
$\approx \pm$ 1\% in $T_{\rm{c}}$,
$\approx \pm$ 3\% in $\rho_{\rm{c}}$,
$\approx \pm$ 1\% in $\xi_{2.5}$, and
$\approx \pm$ 1\% in  $X_{\rm{c}}(^{14}\rm{N})$
for both solar and subsolar models.
The $^{14}$N($p,\gamma$)$^{15}$O reaction rate is the slowest step
in the CNO cycle and thus determines the rate at which H is depleted
in the core \citep[e.g.,][]{iliadis_2007_aa}.
STARLIB currently adopts the reaction rate of \citet{imbriani_2005_aa}.

The lowest positive-energy resonance of
$^{14}$N($p,\gamma$)$^{15}$O is located at a center-of-mass
energy of 259 keV, too high in energy to strongly influence quiescent
stellar burning \citep[e.g.,][]{luna-collaboration_2006_aa}.
However, the strength of this resonance is often used as
a cross-section normalization for lower-energy measurements.
\citet{daigle_2016_aa}
report measurements of the energy, strength, and $\gamma$-ray
branching ratios for the 259 keV resonance. Their recommended
strength of $\omega\gamma$ = 12.6 MeV is in agreement with
the previous value but more precise, and offers a more reliable normalization.
Using this result, they suggest  the $S$-factor data of
\citet{imbriani_2005_aa} should be reduced by 2.3\%.
For this reduction of the $S$-factor, in our stellar models at H-depletion
the largest variation is $\approx$\,+0.2\% with respect to the mean for $\rho_{\rm{c}}$.
Other properties have variations $\lesssim 0.1\%$.

STARLIB currently adopts the triple-$\alpha$ reaction rate of \citet{angulo_1999_aa}.
Uncertainties in this reaction rate dominate the 95\% CI variations of
$\approx \pm$ 1.5\% in $T_{\rm{c}}$,
$\approx \pm$ 3.5\% in $\rho_{\rm{c}}$, and
$\approx~\pm$ 3.5\% in $\xi_{2.5}$,
during core He-burning.
\citet{nguyen_2012_aa} combine Faddeev hyperspherical harmonics and the
$R$-matrix method to suggest the triple-$\alpha$ reaction rate is significantly
enhanced at temperatures below 0.06 GK. For an increased reaction rate in this
temperature range, our analysis suggest
$T_{\rm{c}}$ and $\rho_{\rm{c}}$ will decrease by $\approx$\,2\% in our \MESA \ models.

STARLIB currently adopts the \citet{kunz_2002_aa} reaction rate
for $^{12}$C($\alpha,\gamma$)$^{16}$O. Experimental uncertainties in this reaction rate
dominate the 95\% CI variations of
$\approx$ 2.5\% in $M_{\rm{CO-Core}}$,
$\approx \pm$ 1\% in $\tau_{\rm{He-burn}}$,
$\approx$ +80/-70\% in $X_{\rm{c}}(^{12}$C), and
$\approx$ +25/-27\% in $X_{\rm{c}}(^{16}$O) during core He-burning.
Core C-, Ne-, and even O-burning inherit some of these dependencies.
\citet{deboer_2017_aa} use the $R$-matrix method to derive a new
$^{12}$C($\alpha,\gamma$)$^{16}$O rate. The uncertainties in
the \citet{deboer_2017_aa} rate are smaller
than the uncertainties in the \citet{kunz_2002_aa} rate near temperatures
of $0.05 \lesssim T \lesssim 1 $ GK and slightly larger near $T \simeq 1-3 $ GK.
\citet{deboer_2017_aa} show their rate can lead to changes of $\simeq$\,$\pm$\,1.5\%
for $M_{\rm{CO-Core}}$ in their 15\,\Msun solar metallicity \MESA \ models. 
This is slightly smaller than our 95\% CI spread.

STARLIB adopts the {$^{12}$C\,+$^{12}$C, {$^{12}$C\,+$^{16}$O, and
{$^{16}$O\,+$^{16}$O rates and branching ratios of \citet{caughlan_1988_aa}.
Uncertainties in these reaction rates and branching ration dominate the 95\% CI variations of
$\approx$ +23/-50\% in $M_{\rm{ONe-Core}}$ at C-depletion
and $\approx$\,+40/-35\% in $\rho_{\rm{c}}$ at Ne-depletion.

The {$^{12}$C\,+$^{12}$C is one of the most studied heavy ion reactions.
Despite several decades of dedicated experimental efforts, the low-energy reaction rate
still carries considerable uncertainties due to pronounced resonance structures
that are thought to be associated with molecular configurations of carbon
in the $^{24}$Mg excited state \citep[e.g.,][]{misicu_2007_aa}
However, it has been argued that low-energy cross section of fusion
reactions declines faster with decreasing energy than projected by
common potential models \citep[][]{jiang_2007_aa, gasques_2007_aa, carnelli_2014_aa}.

The impact of changes in the {$^{12}$C\,+$^{12}$C in 1D
Geneva stellar evolution (GENEC) models
are investigated in \citet{bennett_2012_aa} and \citet{pignatari_2013_aa}.
They find that an increase in the {$^{12}$C\,+$^{12}$C reaction 
rate causes core C-burning ignition at lower temperature. This reduces the
thermal neutrino losses, which in turn increases the core C-burning lifetime.
They also find an increased {$^{12}$C\,+$^{12}$C rate increases the upper initial mass
limit for when a star undergoes convective C-burning rather than
radiative C-burning \citep{lamb_1976_aa,woosley_1986_aa, petermann_2017_aa}.  
The subsequent evolution of these more massive stars may yield
a bimodal distribution of compact objects
\citep{timmes_1996_ac,zhang_2008_ab,petermann_2017_aa}.

\citet{fang_2017_aa} use a high-intensity oxygen beam impinging upon
an ultrapure graphite target to make new measurements of the total
cross section and branching ratios for the {$^{12}$C\,+$^{16}$O reaction. They find
a new broad resonance-like structure and a decreasing trend
in the $S$-factor data towards lower energies, in contrast to previous
measurements. For massive stars, they conclude the impact of the new rate 
{$^{12}$C\,+$^{16}$O rate might be small for core and shell burning
\citep[also see][]{jiang_2007_ab}, although the impact might be enhanced by
multidimensional turbulence \citep{cristini_2017_aa}
or rotation \citep{chatzopoulos_2016_aa} of the pre-supernova star
during the last phases of its stellar life.

Of the key nuclear reaction rates identified in this study, those
with the largest uncertainty over the temperature ranges consider here are
heavy ion $^{12}$C\,+$^{12}$C, $^{12}$C\,+$^{16}$O, and $^{16}$O\,+$^{16}$O 
reactions. Due to the larger Coulomb barrier for the $^{12}$C\,+$^{16}$O reaction
it is expected to be less efficient during carbon burning. 
Our results suggest that variation in this rate, especially the $p$ exit channel,
can lead to non-negligible variations in core temperature and density during
carbon burning. Our results suggest that for a decrease in the uncertainty in these 
heavy ion reactions  rates over stellar temperatures, along with the 
$^{12}$C($\alpha,\gamma$)$^{16}$O reaction, we can expect a \emph{decrease} 
in variation of stellar model properties of below the level of variations induced by 
uncertainties due to stellar winds, convective boundary mixing, and 
mass/network resolution.

\subsection{Assessing The Overall Impact}\label{s:assessing}

Paper F16 applies the Monte Carlo framework to stellar
models that form CO white dwarfs. They evolve 3\,\Msun solar
metallicity models from the pre-MS to the first thermal pulse. They
sample 26 out of 405 nuclear reactions and consider one evolutionary epoch $-$
the first thermal pulse, a time shortly after core
He-depletion. Comparing our Figure~\ref{fig:all_rhos_he_dep} with
their Figure~11, we find similar results despite the different masses. The
$^{12}$C($\alpha,\gamma$)$^{16}$O dominates the mass of the CO core,
The $^{12}$C, and $^{16}$O mass fractions at He-depletion (their first
thermal pulse) have similar sign and magnitude SROC coefficients.  In
agreement with their CO white dwarf models, variations in the central
temperature are driven by uncertainties in the triple-$\alpha$
reaction rate. They report that the central density is
primarily correlated with uncertainties in the $^{12}$C($\alpha,\gamma$)$^{16}$O rate,
while we find the variations in the central density are chiefly correlated
with uncertainties in the triple-$\alpha$ rate. This difference is due to the masses considered.
The hotter, less dense, cores of our 15\,\Msun models favor the
triple-$\alpha$ rate as the primarily source of the central density variations,
whereas the cooler, more dense 3\,\Msun models favor $^{12}$C($\alpha,\gamma$)$^{16}$O.

\citet{farmer_2016_aa} explore uncertainties in the structure of massive star
stellar models with respect to mass resolution, mass loss, and the number of
isotopes in the nuclear reaction network. \citet{farmer_2016_aa} and this paper both 
report results for 15\,\Msun, $\dot{M}$\,$\ne$\,0, 127 isotope, solar metallicity, 
\MESA r7624 models. The primary difference between this paper and 
\citet{farmer_2016_aa} is the use of STARLIB reaction rates.

Our results at H-depletion can be compared with their results at He-ignition.
For example, Table \ref{tbl:all_properties} shows our mean He core mass is 
$M_{\rm{He-Core}}$\,=\,2.80\,\Msun while their median He core mass is 
He$_{\rm core}$\,=\,2.77\,\Msun, a difference of $<$\,1\%. Our 95\% CI for 
$M_{\rm{He-Core}}$ is within 1\% of their He$_{\rm core}$ upper and
lower limits. As another example, our mean H burning lifetime is 
$\tau_{\rm{TAMS}}$\,=\,11.27\,Myr and their median H burning lifetime is 
$\tau_{\rm{H}}$\,=\,10.99 Myr, a difference of $\simeq$\,3\%.  In
addition, our 95\% CI for $\tau_{\rm{TAMS}}$ is $\simeq$\,2\% larger
than their upper and lower bounds for $\tau_{\rm{H}}$.

Our He-depletion results can also be compared to their results at C-ignition.
We find a mean $M_{\rm{C-Core}}$\,=\,2.41\,\Msun while
their median C$_{\rm core}$\,=\,2.44\,\Msun, a difference of $<$ 1\%.
Our 95\% CI spread due to uncertainties in the nuclear reaction rates
is $\simeq$\,(+1.9\%, $-$3.1\%) while their upper and lower bounds
suggest variations of $\simeq$\,(+3.7\%, $-$0.4\%) due to changes in
mass and network resolution.  In addition, our mean
$\tau_{\rm{He-burn}}$\,=\,1.594\,Myr and their median
$\tau_{\rm{He}}$\,=\,1.74 Myr, a difference of $\simeq$\,8\%.  Our 95\%
CI for $\tau_{\rm{He-burn}}$ is $\simeq$\,(+1.9\%, $-$3.1\%) while their
upper and lower bounds are $\simeq$\,(+1.2\%, $-$12.1\%).

Comparing our Ne-depletion results with their O-ignition results,
we find a mean $M_{\rm{O-Core}}$\,=\,1.44\,\Msun while
their median O$_{\rm core}$\,=\,1.40\,\Msun, a difference of $\lesssim$\,1\%.
Our 95\% CI spread due to uncertainties in the nuclear reaction rates
is $\simeq$\,(+65\%, $-$23\%) while their upper and lower bounds
suggest variations of $\simeq$\,(+0.1\%, $-$5.6\%) due to changes in
mass and network resolution.  In addition, our mean
$\tau_{\rm{C-burn}}$\,=\,30.74\,kyr and their median
$\tau_{\rm{C}}$\,=\,85.55 yr differs by approximately three orders of
magnitude. This large difference is due to the exact measurement points.
In this work, we assumed the time to be the difference between the age of the
star at C-depletion and He-depletion. This does not necessarily correspond to the exact
burning lifetime for C as the star undergoes reconfiguration after He-depletion 
for a few thousand years before conditions for C-burning are met. In 
\citet{farmer_2016_aa} they measure the time to transition to the next major
fuel source. Our 95\%CI for $\tau_{\rm{C-burn}}$ is $\simeq$\,(+1.9\%, $-$3.1\%) 
while their upper and lower bounds are $\simeq$\,(+1.2\%, $-$12.1\%).

Variations in properties of stellar evolution models can be found to be caused by other 
sources of uncertainty beyond those discussed above. \citet{renzo_2017_aa} considered 
uncertainties in the mass loss prescriptions and efficiencies used in solar-metallicity, 
non-rotating, single stars. They find that  changes in these parameters can lead to a 
spread of $\Delta M_{\rm{CO}}\approx0.28$\,$M_{\odot}$ in CO core masses measured at 
O-depletion, though defined differently in their work as the
moment when $X_{\rm{c}}(^{16}\rm{O})\lesssim 0.04$. This spread represents a 
variation of about $ \pm 5 \%$ variation about the arithmetic mean.
The treatment of mixing at the convective boundaries can also have a significant effect on 
the evolution of massive stellar models. \citet{davis_2018_aa} show that for their 
25\,$M_{\odot}$ model at Ne-ignition, they find a variation of $+ 5\%$ in the ONe core 
mass due to changes in the efficiency of convective boundary mixing at metal burning interfaces. 

\citet{farmer_2016_aa} find that mass resolution has a larger impact
on the variations than the number of isotopes up to and including C
burning, while the number of isotopes plays a more significant role in
determining the span of the variations for Ne-, O-, and Si-burning. 
Comparisons of the core masses and burning lifetimes suggests that at
H- and He-depletion, the variations induced by uncertainties in
nuclear reaction rates are of comparable magnitude to the variations
induced by the modeling choices of mass resolution and network resolution.
At Ne-depletion the integrated impact of the uncertainties in the reaction rates
appear to be larger than the variations caused by mass and network resolution.

The scale of variations due to different mass loss prescriptions and efficiencies 
were found to be of comparable scale to those due to reaction rate uncertainties at 
early epochs such as H- and He-depletion for the stellar properties considered. 
At early epochs, convective boundary mixing is likely to cause significant variations in 
core masses and lifetimes that are of larger scale than those due to nuclear reaction rate 
uncertainties. However, uncertainties in convective boundary mixing are likely to be smaller
than the integrated impact of rate uncertainties at advanced burning stages.

\section{Summary}
\label{sec:summary}

We investigated properties of pre-supernova massive stars with respect to 
the composite uncertainties in the thermonuclear reaction rates by coupling 
the reaction rate PDFs provided by the STARLIB reaction rate library with 
\MESA stellar models. We evolved 1000 15\,\Msun models with solar 
and subsolar initial compositions from the pre main-sequence to core 
oxygen depletion for a total of 2000 Monte Carlo stellar models.
For each stellar model we  sampled 665 forward thermonuclear
reaction rates concurrently, and used them in an in-situ 127 isotope 
\MESA \ reaction network. With this infrastructure we surveyed the core mass, 
burning lifetime, central temperature, central density, compactness parameter, 
and key abundances at H-, He-, C-, Ne-, and O-depletion. 

At each stage, we 
measured the PDFs of the variations of each property and calculated SROC 
coefficients for each sampled reaction rate. This allowed identification of the
reaction rates that have the largest impact on the variations of the properties 
surveyed. Table~\ref{tbl:all_properties} summarizes the stellar properties, 
the reaction rates causing their variation, and the largest correlation coefficient 
(positive or negative) for that reaction rate. 

In general, variations induced by nuclear reaction rates 
grow with each passing phase of evolution.
Relative to variations induced by 
mass resolution and the number of isotopes in the nuclear reaction network,
we found that variations induced by uncertainties in nuclear reaction rates at core
H- and He-depletion are of comparable magnitude to the variations 
induced by the modeling choices of mass resolution and network resolution.
Beyond these evolutionary epochs, our models suggest that the reaction
rate uncertainties can dominate the variation in properties of the stellar model
significantly altering the evolution towards iron core-collapse.

\acknowledgements
This project was supported by NSF under the {\it Software Infrastructure
for Sustained Innovation} (SI2) program grants (ACI-1339581,
ACI-1339600, ACI- 1339606, ACI-1663684, ACI-1663688, ACI-1663696) and
grant PHY-1430152 for the {\it Physics Frontier Center} ``Joint
Institute for Nuclear Astrophysics - Center for the Evolution of the
Elements'' (JINA-CEE).  This project was also supported by NASA under
the {\it Theoretical and Computational Astrophysics Networks} (TCAN) program
grants (NNX14AB53G, NNX14AB55G, NNX12AC72G) and the Astrophysics
Theory Program grant 14-ATP14-0007.
C.E.F. acknowledges support from a Predoctoral Fellowship
administered by the National Academies of Sciences, Engineering,
and Medicine on behalf of the Ford Foundation, an Edward J. Petry
Graduate Fellowship from Michigan State University, and the National
Science Foundation Graduate Research Fellowship Program
under grant number DGE1424871.
S.M.C. is supported by the U.S. Department of Energy, Office of Science, Office of Nuclear Physics, under Award Numbers DE-SC0015904 and DE-SC0017955, the Research Corporation for Science Advancement under Scialog Grant 23770, and the Chandra X-ray Observatory under grant TM7-18005X.
The simulations presented in this work were performed on the \texttt{laconia}
cluster supported by the Institute for Cyber-Enabled Research (ICER) at
Michigan State University.
This research made extensive use of the SAO/NASA Astrophysics Data System (ADS).

\software{
\texttt{MESA} \citep{paxton_2011_aa,paxton_2013_aa,paxton_2015_aa},
STARLIB \citep[][\url{http://starlib.physics.unc.edu}]{sallaska_2013_aa,iliadis_2015_aa,iliadis_2016_aa},
\texttt{Python} available from \href{https://www.python.org}{python.org},
\texttt{matplotlib} \citep{hunter_2007_aa},
\texttt{NumPy} \citep{der_walt_2011_aa}, and
\texttt{scipy} \citep{jones_2001_aa}.}

\bibliographystyle{aasjournal}
\bibliography{msmc}

\end{document}